\documentclass[twocolumn,showpacs,amsfonts,amsmath,amssymb,floatfix,longbibliography]{revtex4-1}
\usepackage{url}
\usepackage{bm}		
\usepackage{graphicx}	
\usepackage{hyperref}

\DeclareMathOperator{\wedgie}{\wedge}

\begin{document}


\title[BKL collapse inside rotating black holes]{Mass inflation followed by Belinskii-Khalatnikov-Lifshitz collapse inside accreting, rotating black holes}

\author{Andrew J S Hamilton}
\email{Andrew.Hamilton@colorado.edu}	
\affiliation{JILA, Box 440, U. Colorado, Boulder, CO 80309, USA}
\affiliation{Dept.\ Astrophysical \& Planetary Sciences,
U. Colorado, Boulder, CO 80309, USA}

\newcommand{\simpropto}{\raisebox{-0.7ex}[1.5ex][0ex]{$\,
                \begin{array}[b]{@{}c@{\;}} \propto \\
                [-1.5ex] \sim \end{array}$}}

\newcommand{\dd}{d}
\newcommand{\dext}{{\rm d}}
\newcommand{\ddi}[1]{\dd^{#1}\mkern-1.5mu}	
\newcommand{\bddi}[1]{\bdd^{#1}\mkern-1.5mu}
\newcommand{\DD}{D}
\newcommand{\ee}{e}
\newcommand{\im}{i}
\newcommand{\Ei}{{\rm Ei}}
\newcommand{\perpperp}{\perp\!\!\perp}
\newcommand{\ppartial}{\partial^2\mkern-1mu}
\newcommand{\nn}{\nonumber\\}
\newcommand{\transpose}{\top}

\newcommand{\diag}{{\rm diag}}
\newcommand{\jel}{\text{\sl j}}
\newcommand{\Lz}{L}
\newcommand{\Msun}{{\rm M}_\odot}
\newcommand{\uel}{u}
\newcommand{\vel}{v}
\newcommand{\inn}{{\rm in}}
\newcommand{\out}{{\rm ou}}
\newcommand{\sep}{{\rm sep}}

\newcommand{\be}{\bm{e}}
\newcommand{\bg}{\bm{g}}
\newcommand{\bp}{\bm{p}}
\newcommand{\bpi}{\bm{\pi}}
\newcommand{\bPi}{\bm{\Pi}}
\newcommand{\bR}{\bm{R}}
\newcommand{\bS}{\bm{S}}
\newcommand{\bSpin}{\bm{\Sigma}}
\newcommand{\bT}{\bm{T}}
\newcommand{\bvartheta}{\bm{\vartheta}}
\newcommand{\bu}{\bm{u}}
\newcommand{\bv}{\bm{v}}
\newcommand{\bx}{\bm{x}}
\newcommand{\bgamma}{\bm{\gamma}}
\newcommand{\bGamma}{\bm{\Gamma}}
\newcommand{\bpartial}{\bm{\partial}}

\newcommand{\KCarter}{{\cal K}}
\newcommand{\Mass}{{\cal M}}
\newcommand{\mbh}{m_\bullet}
\newcommand{\Mbh}{M_\bullet}
\newcommand{\Mbhdot}{\dot{M}_\bullet}
\newcommand{\NUT}{{\cal N}}
\newcommand{\Qelec}{Q}
\newcommand{\Qelecbh}{\Qelec_\bullet}
\newcommand{\Qmag}{{\cal Q}}
\newcommand{\Qmagbh}{\Qmag_\bullet}
\newcommand{\rhosep}{\rho_{\rm s}}
\newcommand{\tform}{\bar{t}}
\newcommand{\alphaform}{\bar{\alpha}}
\newcommand{\unit}[1]{\, {\rm{#1}}}
\newcommand{\xin}{x_{\rm in}}

\newcommand{\kninfBHfourtwoafig}{
    \begin{figure*}[tb!]
    \begin{center}
    \leavevmode
    \includegraphics[scale=.67]{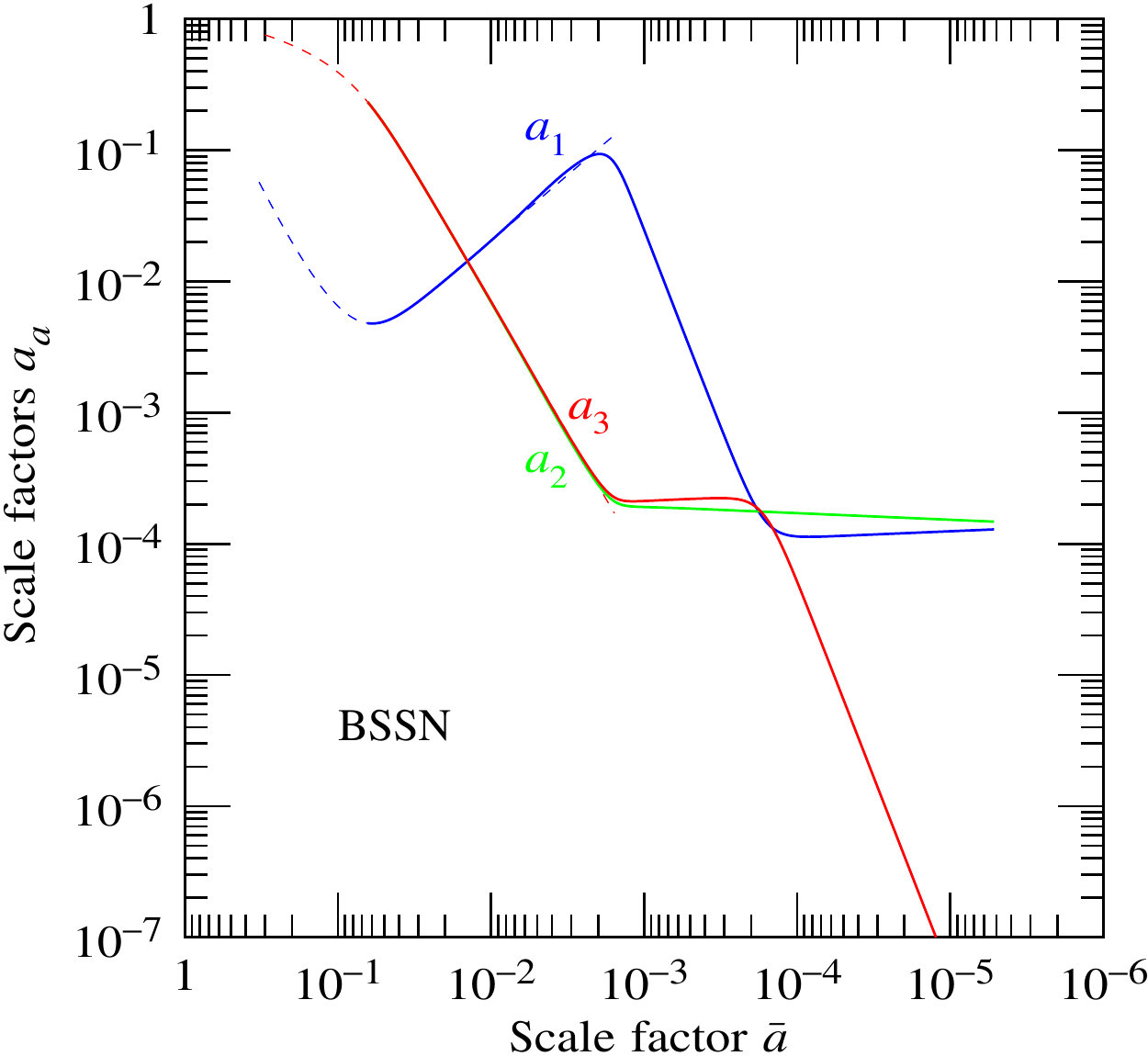}
    \hskip2em
    \includegraphics[scale=.67]{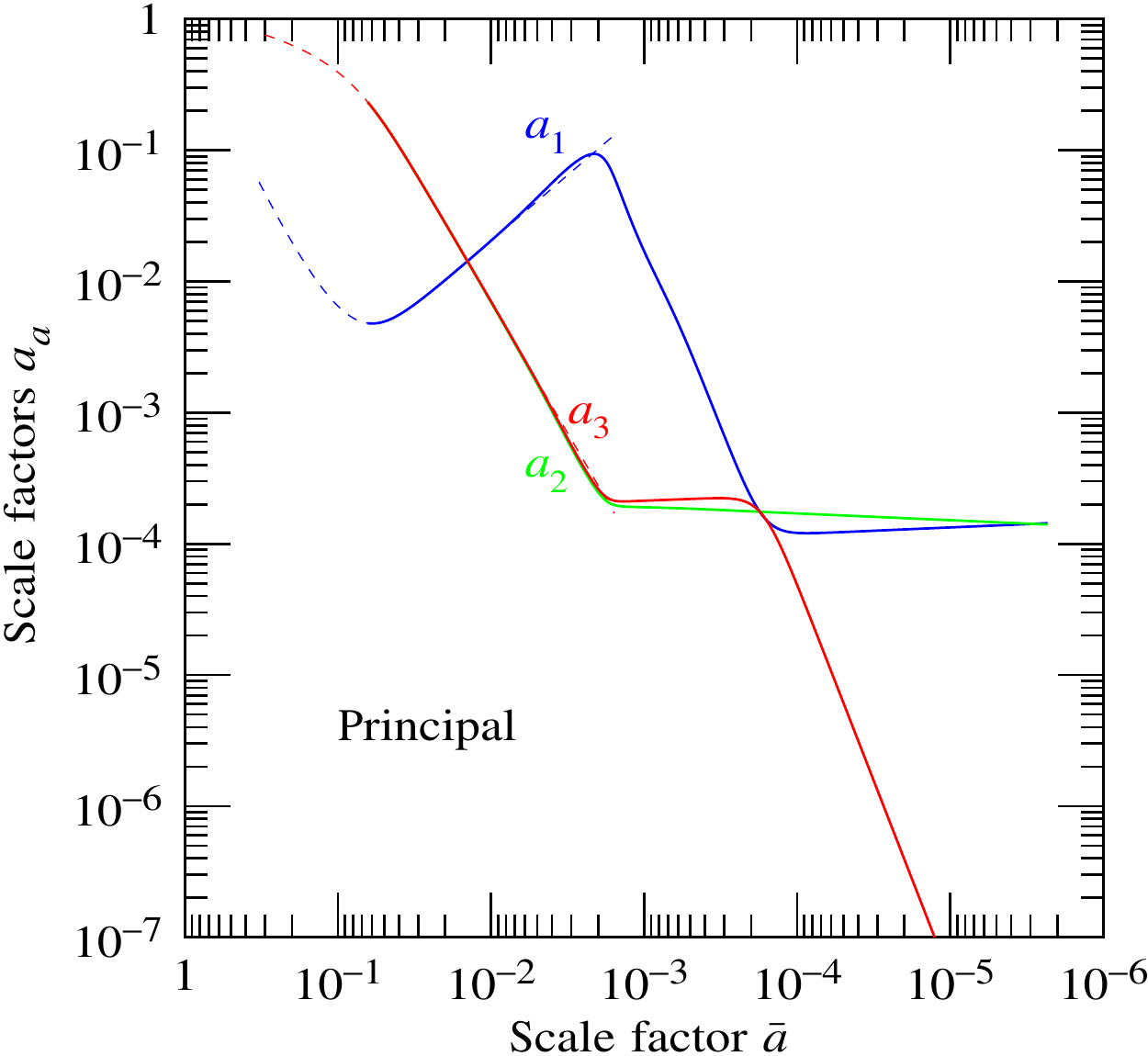}
    \vskip2ex
    \includegraphics[scale=.67]{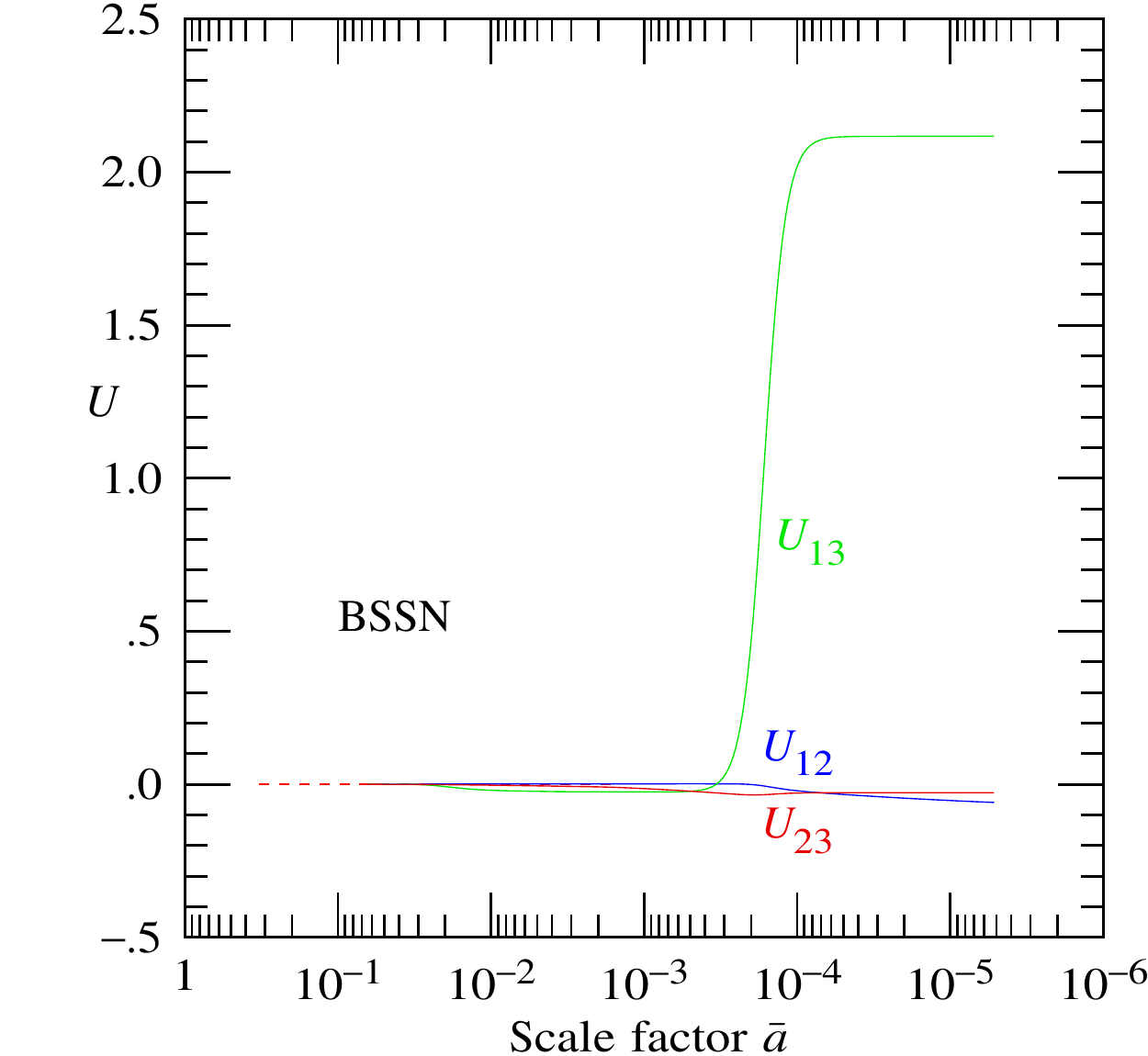}
    \hskip2em
    \includegraphics[scale=.67]{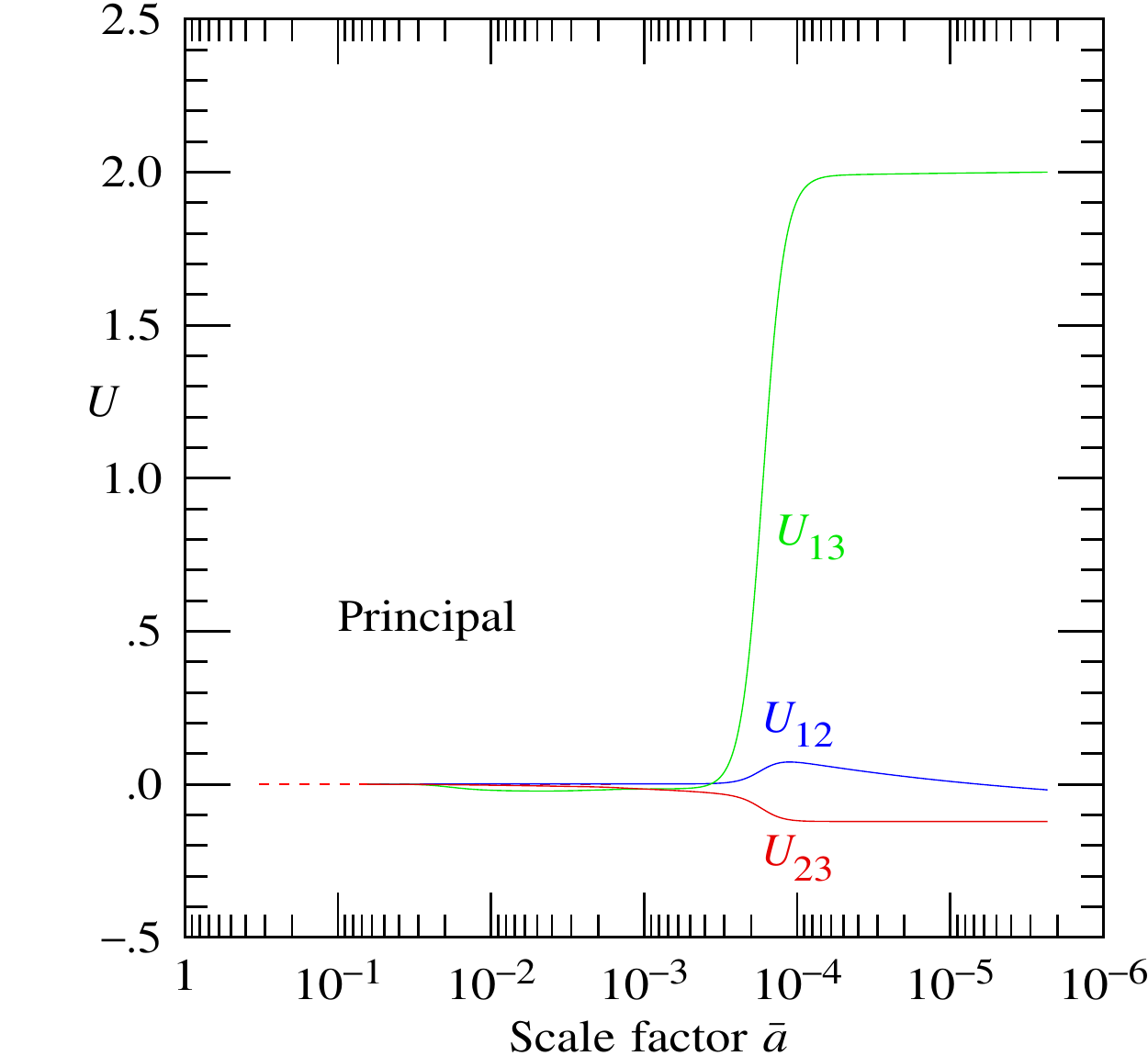}
    \caption{
    \label{kninfBH42a}
Evolution of (top) the lengths $a_a$,
and (bottom) spatial rotation $U_{ab}$,
equations~(\ref{aUdyn}),
of the three axes
of the spatial vierbein during BKL collapse
inside an accreting, rotating black hole,
as a function of the mean scale factor $\bar{a} \equiv ( a_1 a_2 a_3 )^{1/3}$
(the cube root of the spatial volume element),
along a trajectory into the black hole at $42^\circ$ latitude.
The left and right panels are for the two gauges considered in this paper,
Principal and BSSN, \S\ref{gauge-sec}.
The accretion rates and spin of the black hole are given
by equations~(\ref{vuparameters}) and (\ref{aparameter}).
The dashed lines are the approximate conformally separable solution from
\cite{Hamilton:2010a,Hamilton:2010b},
while the solid lines are from the numerical computation.
The behavior is characteristic of BKL collapse:
there are always two collapsing axes and one expanding axis.
Power-law Kasner epochs are punctuated by BKL bounces
in which one of the collapsing axes turns into expansion,
and the previously expanding axis turns into collapse.
    }
    \end{center}
    \end{figure*}
}

\newcommand{\kninfBHfourtwoqfig}{
    \begin{figure*}[tb!]
    \begin{center}
    \leavevmode
    \includegraphics[scale=.67]{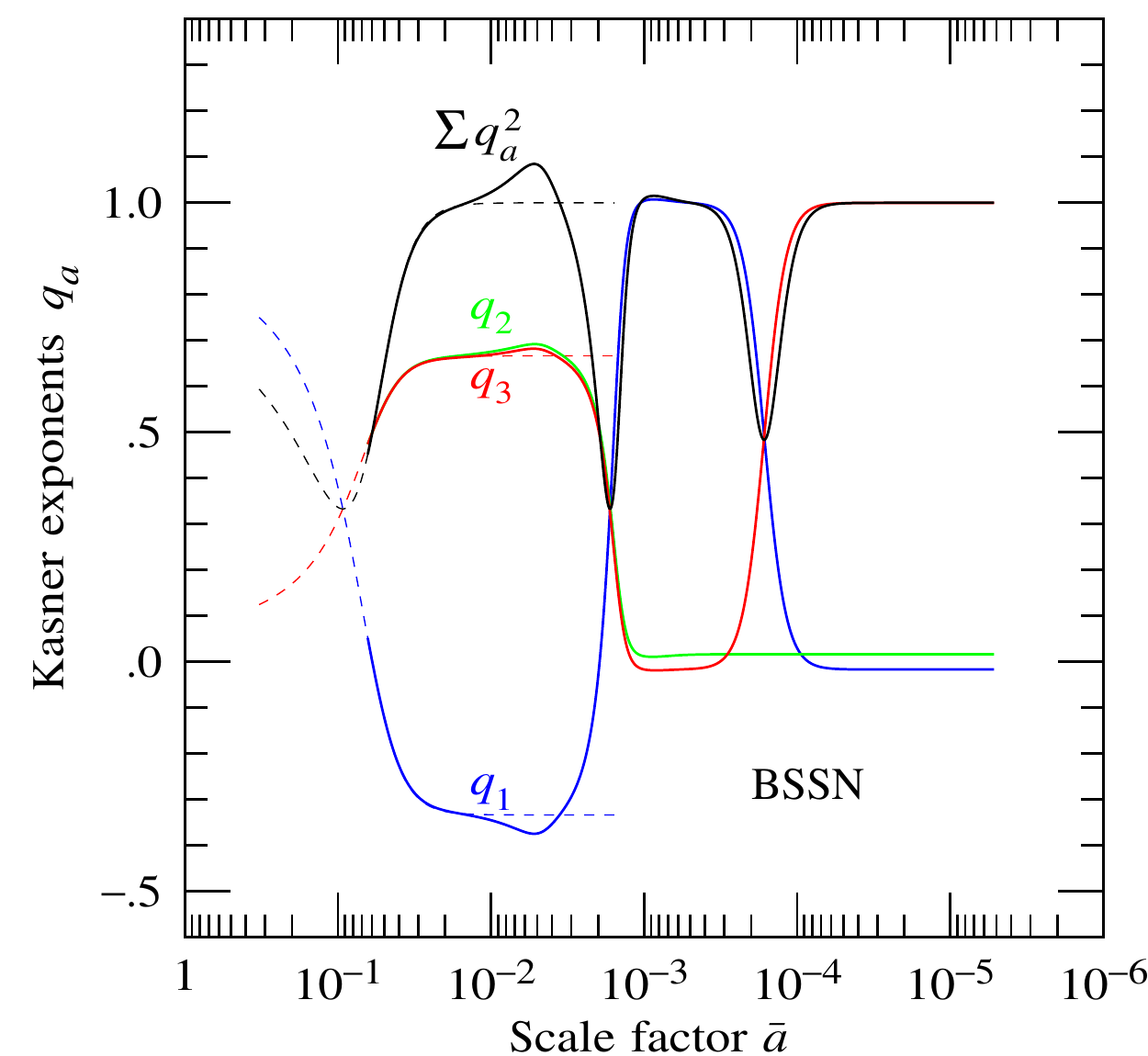}
    \hskip2em
    \includegraphics[scale=.67]{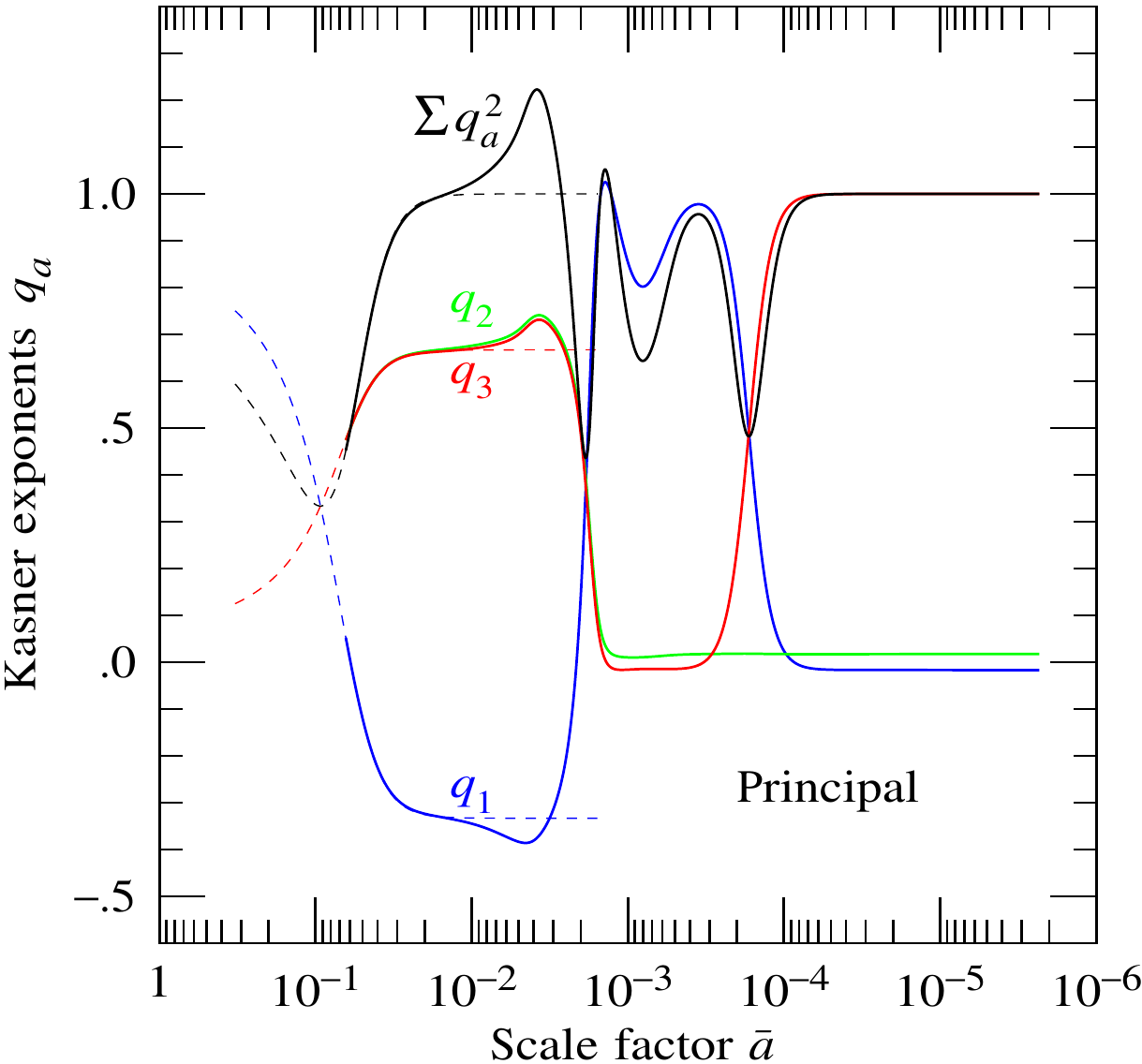}
    \caption{
    \label{kninfBH42q}
Evolution of the Kasner exponents $q_a$
in the same model as Figure~\ref{kninfBH42a},
for BSSN and principal gauges.
    }
    \end{center}
    \end{figure*}
}

\newcommand{\kninfBHfourtwoHfig}{
    \begin{figure*}[tb!]
    \begin{center}
    \leavevmode
    \includegraphics[scale=.67]{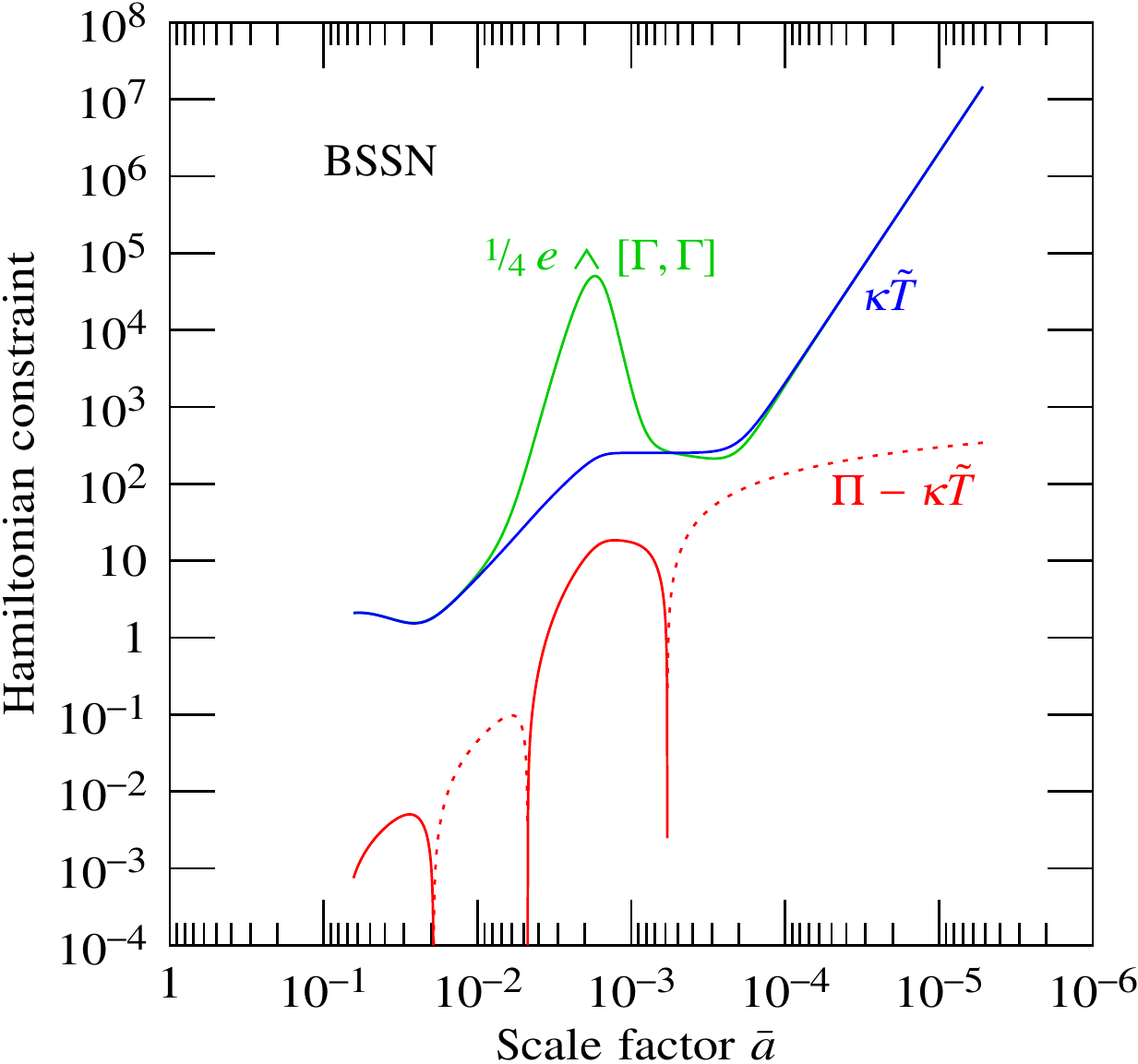}
    \hskip2em
    \includegraphics[scale=.67]{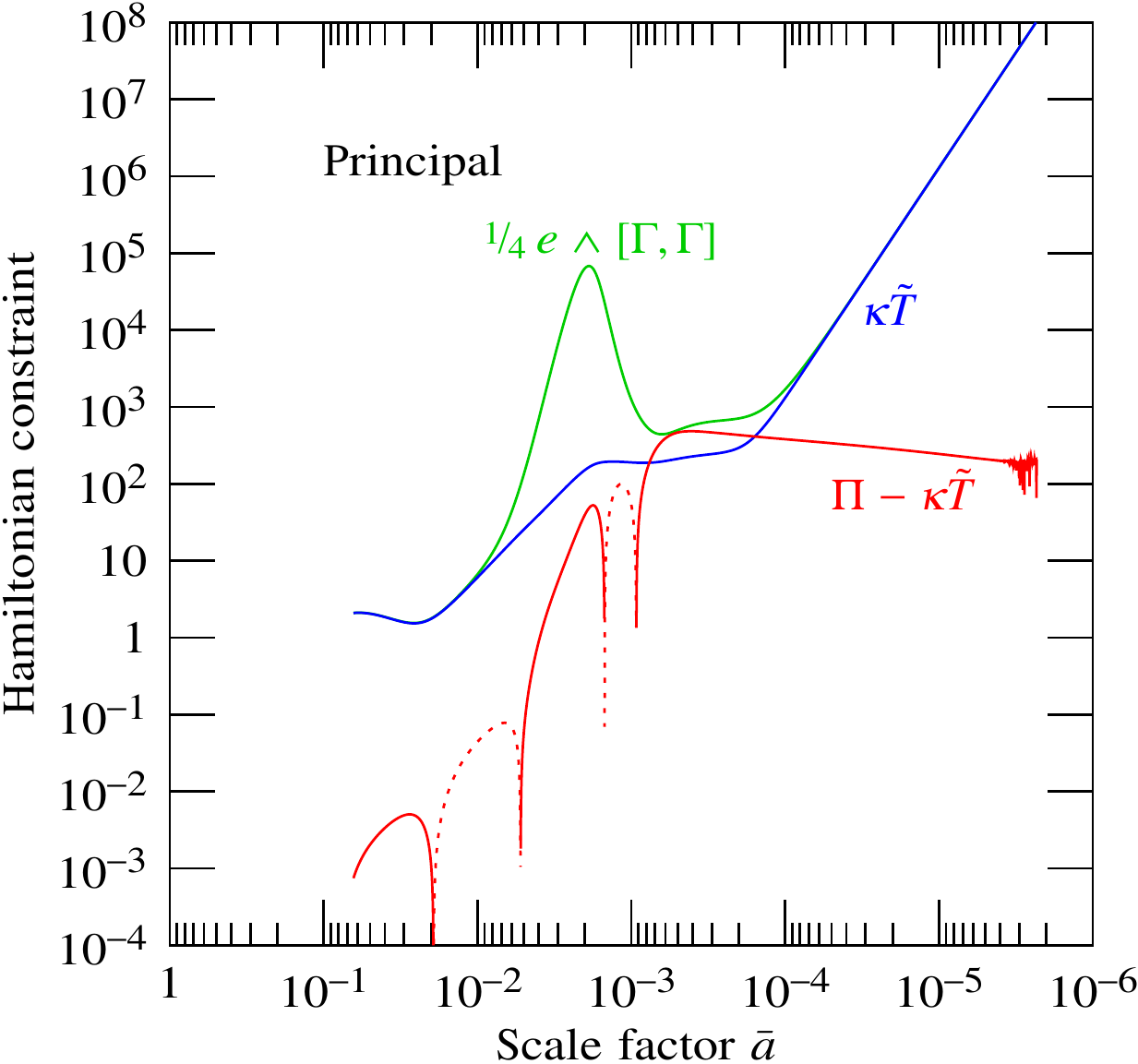}
    \caption{
    \label{kninfBH42H}
Hamiltonian constraint
$( \Pi - \kappa \tilde{T} )_{123\alpha\beta\gamma}$,
equation~(\ref{Ham}),
in the same model as Figure~\ref{kninfBH42a}.
Shown for comparison are the (modified) energy-momentum
tensor $\tilde{T}_{123\alpha\beta\gamma}$
and the potential energy term
$\left( \tfrac{1}{4} \be \wedgie [ \bGamma , \bGamma ] \right)_{123\alpha\beta\gamma}$
in equation~(\ref{Einsteineqaltformd}).
Lines are short-dashed where values are negative.
    }
    \end{center}
    \end{figure*}
}

\newcommand{\kninfBfourtwoDrfig}{
    \begin{figure*}[tb!]
    \begin{center}
    \leavevmode
    \includegraphics[scale=.67]{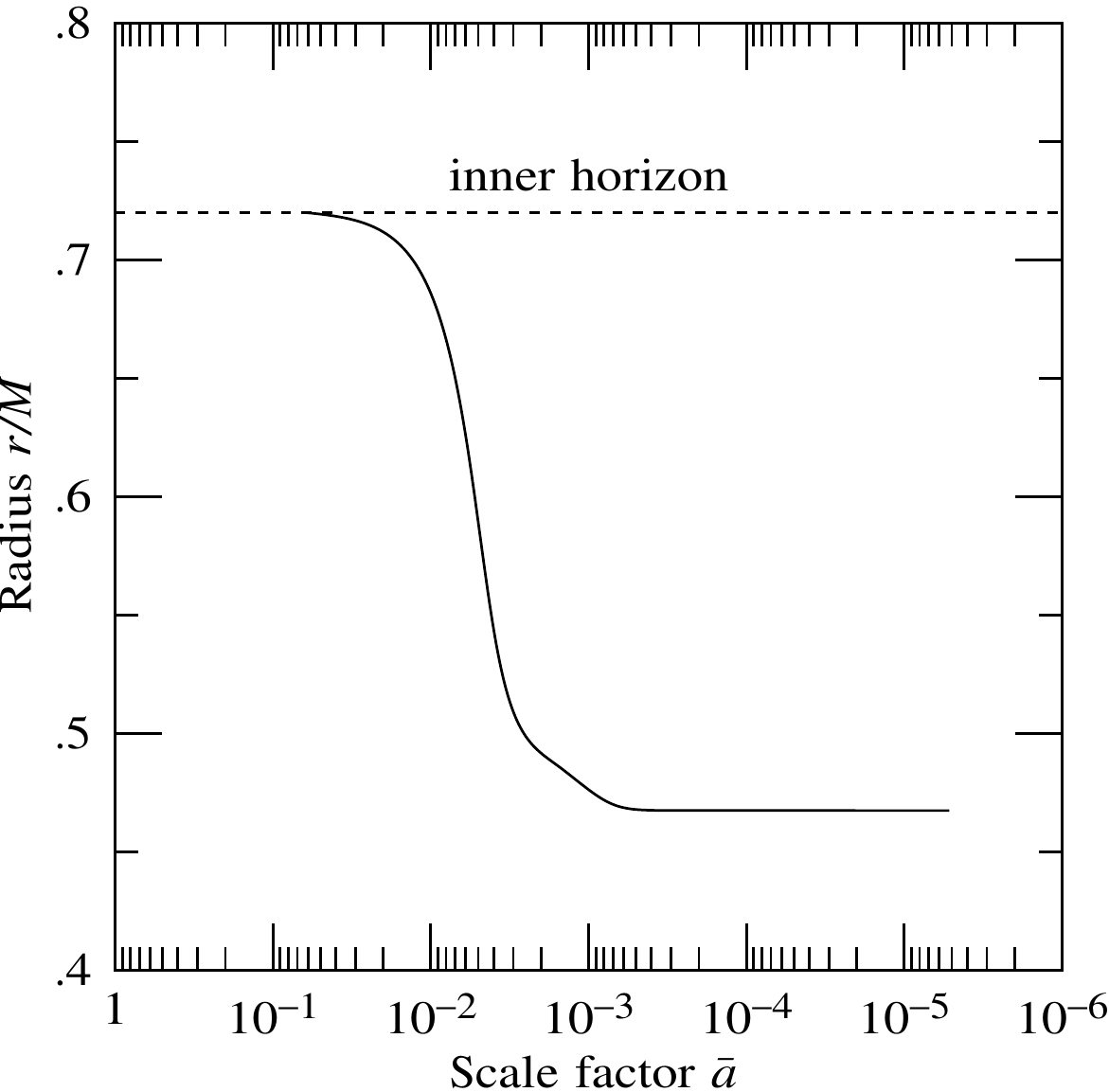}
    \hskip2em
    \includegraphics[scale=.67]{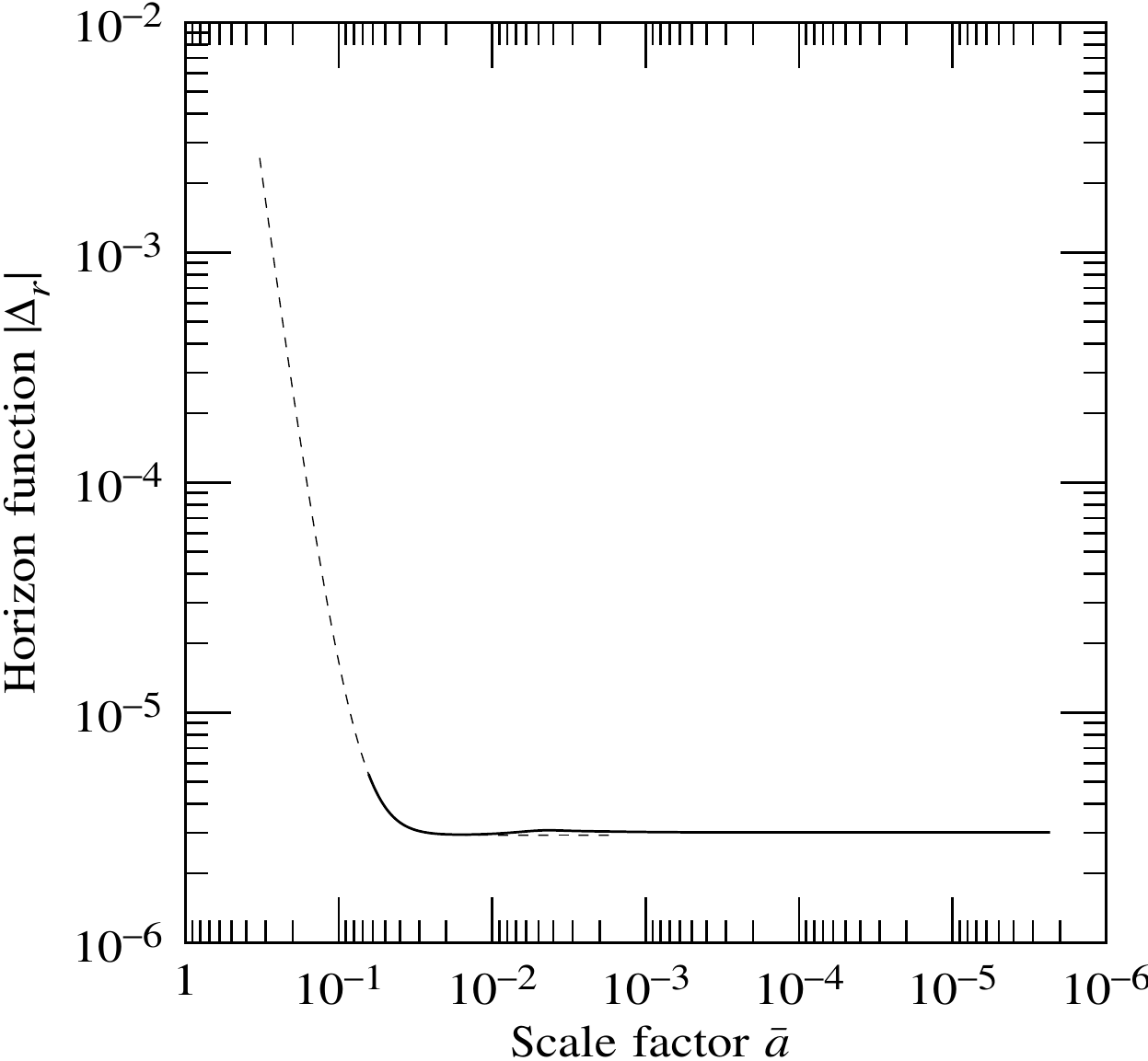}
    \caption{
    \label{kninfB42Dr}
(Left) Conformal radius $r$
in units of the conformal black hole mass $M$,
and
(right)
horizon function $\Delta_r$
(which is negative),
in the model of Figure~\ref{kninfBH42a},
in BSSN gauge.
The results in principal gauge are practically indistinguishable
from those in BSSN gauge.
    }
    \end{center}
    \end{figure*}
}

\newcommand{\kninfBHfourtwomomfig}{
    \begin{figure*}[tbp!]
    \begin{center}
    \leavevmode
    \includegraphics[scale=.67]{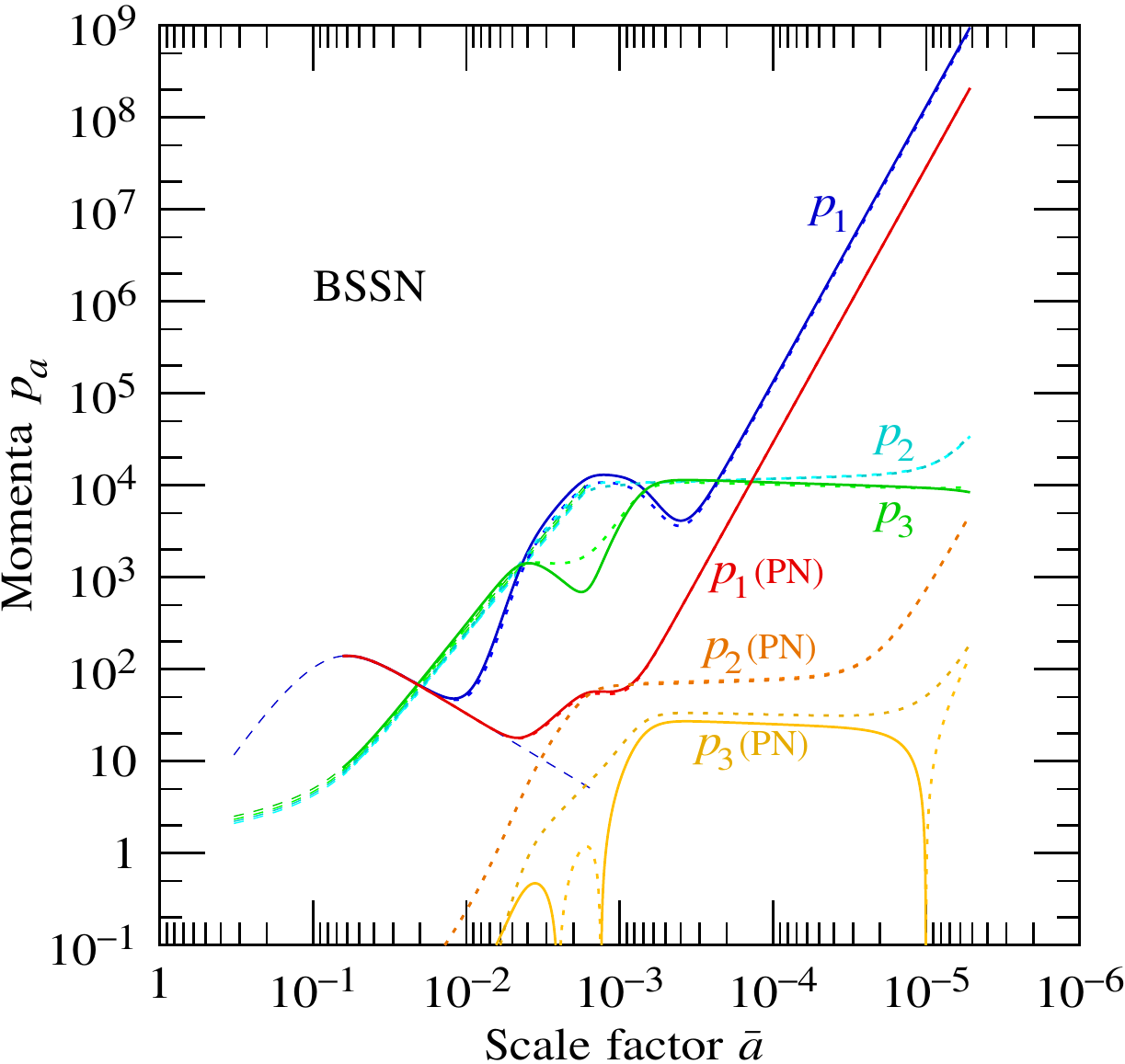}
    \includegraphics[scale=.67]{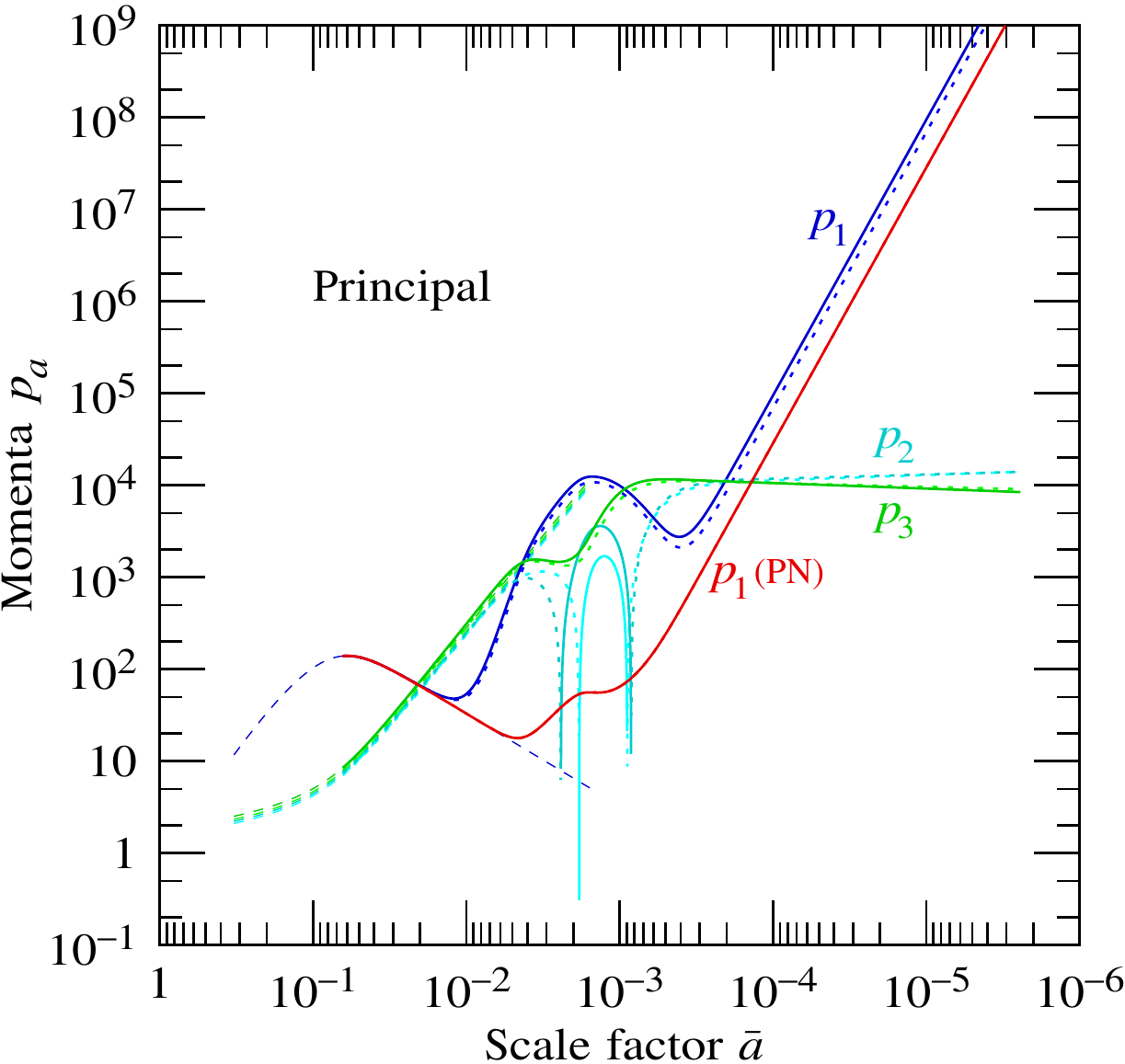}
    \includegraphics[scale=.67]{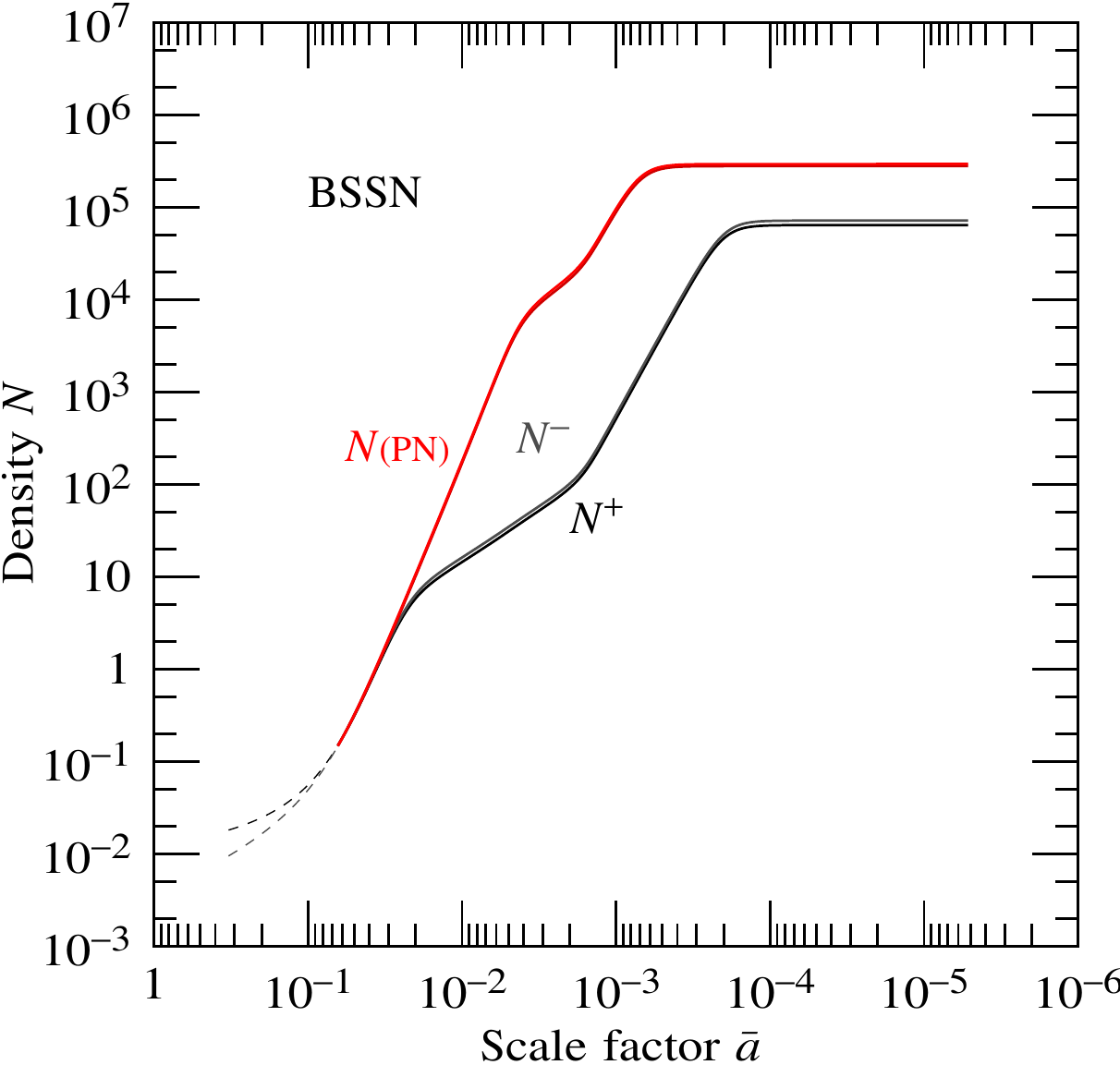}
    \includegraphics[scale=.67]{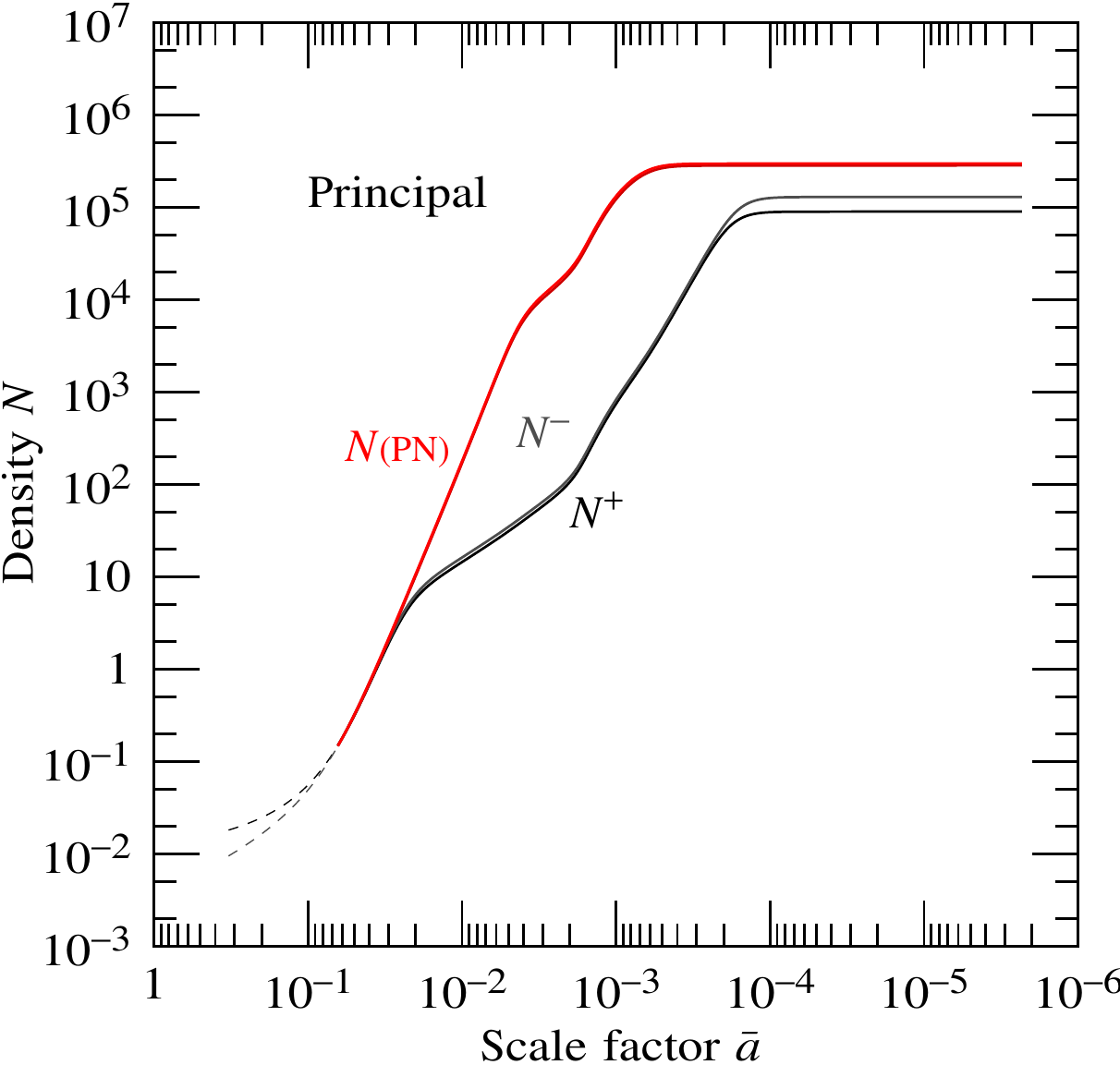}
    \caption{
    \label{kninfBH42mom}
Spatial components $p^\pm_a$ of tetrad-frame momenta,
and densities $N^\pm$,
of outgoing and ingoing collisionless null streams,
in the same model as Figure~\ref{kninfBH42a}.
Lines are short-dashed where values are negative.
Momenta and densities of principal outgoing and ingoing null streams
are also shown, designated {\scriptsize{(PN)}}.
By construction,
in principal gauge,
the principal outgoing and ingoing directions
lie along the tetrad radial direction (the 1-direction).
    }
    \end{center}
    \end{figure*}
}

\newcommand{\kninfBafig}{
    \begin{figure*}[tbp!]
    \begin{center}
    \leavevmode
    \includegraphics[scale=1]{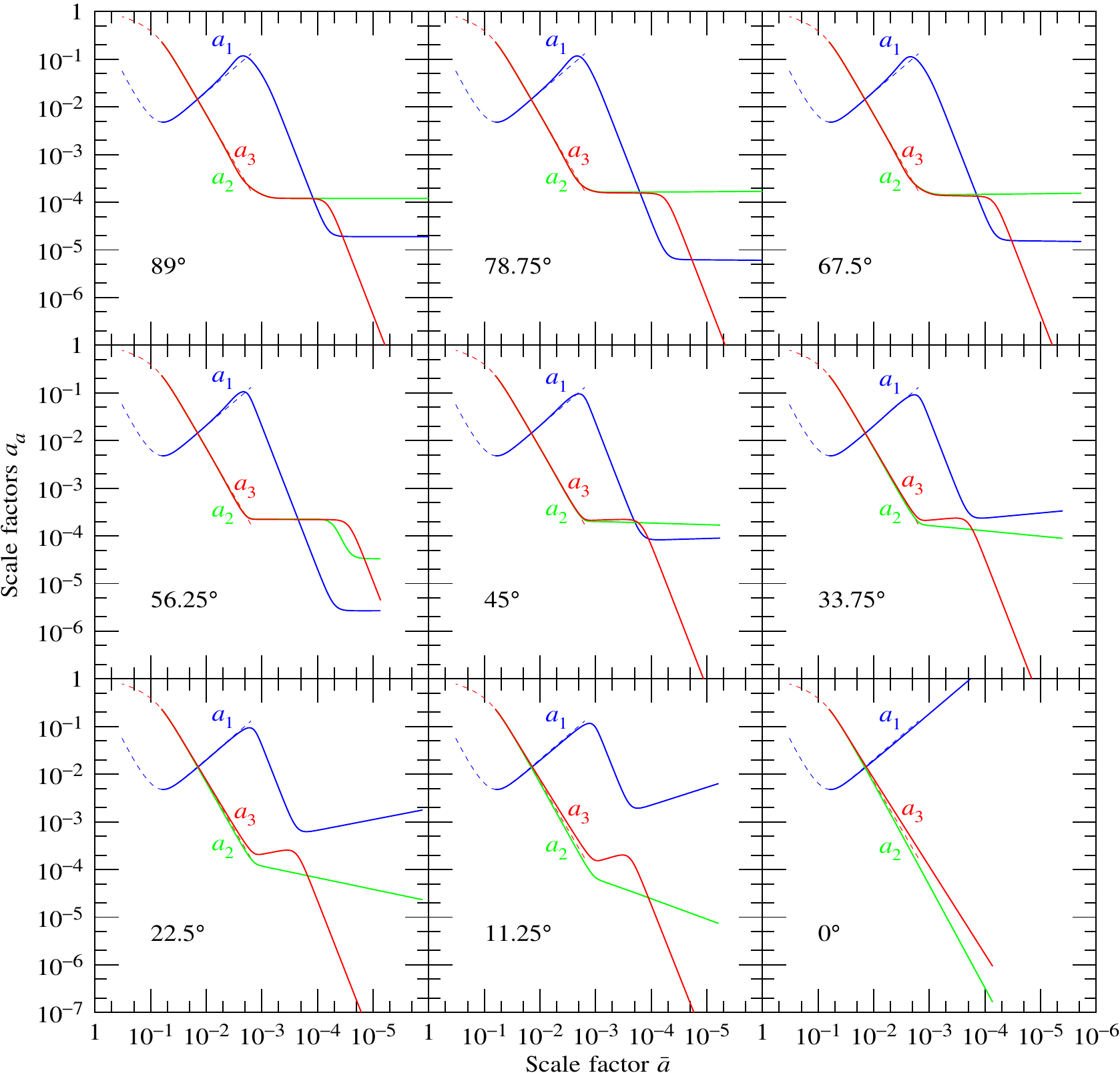}
    \caption{
    \label{kninfBa}
Evolution of the scale factors $a_a$
at various latitudes,
in BSSN gauge.
The accretion rates and spin of the black hole are given
by equations~(\ref{vuparameters}) and (\ref{aparameter}).
The dashed lines are the approximate conformally separable solution from
\cite{Hamilton:2010a,Hamilton:2010b},
while the solid lines are from the numerical computation.
    }
    \end{center}
    \end{figure*}
}

\newcommand{\kninfBqfig}{
    \begin{figure*}[tbp!]
    \begin{center}
    \leavevmode
    \includegraphics[scale=1]{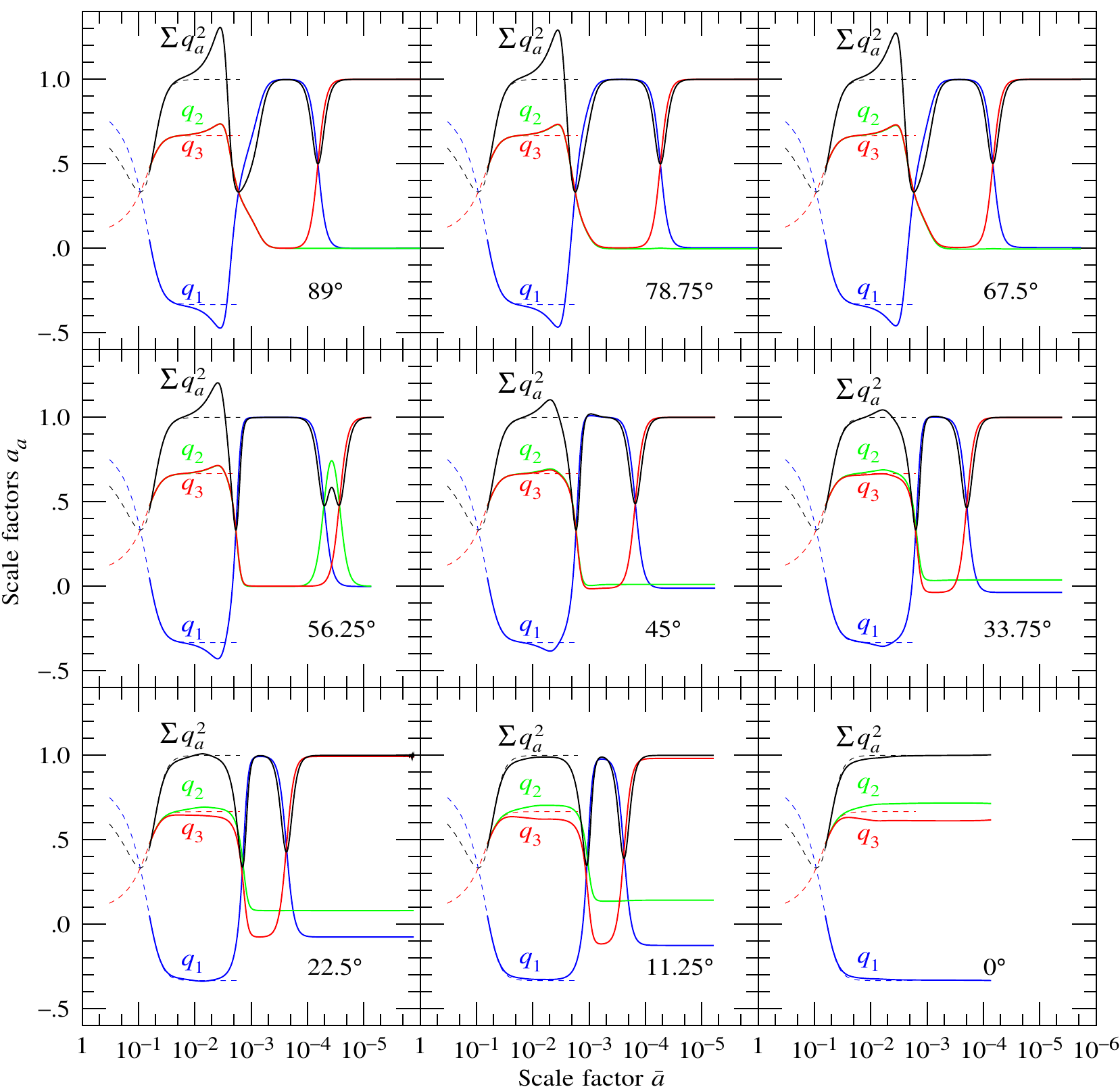}
    \caption{
    \label{kninfBq}
Evolution of the Kasner exponents $q_a$,
for the same models as in Figure~\ref{kninfBa},
again in BSSN gauge.
    }
    \end{center}
    \end{figure*}
}


\hyphenpenalty=3000

\begin{abstract}
Numerical evidence is presented that the Poisson-Israel mass inflation instability at the inner horizon of an accreting, rotating black hole is generically followed by Belinskii-Khalatnikov-Lifshitz oscillatory collapse to a spacelike singularity.
The computation involves following all 6 degrees of freedom of the
gravitational field.
To simplify the problem,
the computation takes as initial conditions
the conformally separable solutions of
\cite{Hamilton:2010a,Hamilton:2010b}
just above the inner horizon of a slowly accreting, rotating black hole,
and integrates the equations inward along single latitudes.
\end{abstract}


\date{\today}

\maketitle

\section{Introduction}


During the 1970s,
Belinskii, Khalatnikov, and Lifshitz
\cite{Belinskii:1970,Belinskii:1971,Belinskii:1972,Belinskii:1982}
(BKL) developed arguments that the generic outcome
of general relativistic collapse to a spacelike singularity
would be complicated and oscillatory.
BKL's arguments were asymptotic in nature.
They assumed that the spacetime was already deep into collapse
towards a spacelike singularity,
and they analyzed the behavior in the asymptotic limit
as the spacetime approached the singularity.

The BKL scenario was brought into question
\cite{Ori:1991,Ori:1992,Ori:2001pc}
by Poisson \& Israel's
\cite{Poisson:1989zz,Poisson:1990eh,Barrabes:1990}
discovery of the mass inflation instability
at the inner horizons of black holes.
Mass inflation is the nonlinear consequence
of the infinite blueshift at the inner horizon first pointed out
by Penrose \cite{Penrose:1968}.
Poisson \& Israel
argued that cross flow between outgoing and ingoing streams
just above the inner horizon would drive exponential growth
of the interior mass.

It has been commonly but incorrectly asserted
that the generic outcome of inflation is
a weak null singularity on the Cauchy horizon
\cite{Dafermos:2012ny,Luk:2013cqa},
not a spacelike singularity.
The conclusion that the singularity is weak and null
is premised on a black hole that collapses and thereafter remains isolated,
in which case the outgoing and ingoing streams that drive inflation
are provided by Price tails of gravitational radiation
generated during collapse,
as originally suggested by
\cite{Poisson:1990eh}.
However, this premise never holds for an astronomical black hole.
The energy density in the most slowly decaying mode (the quadrupole)
of Price radiation decays as $t^{-12}$.
It is straightforward to estimate that the energy density of a Price tail
in the collapse of a stellar-mass black hole
falls below the energy density of, for example, the cosmic microwave background
in of the order of $10^4$ black hole crossing times, or of order 1 second.

As argued by \cite{Hamilton:2008zz},
during inflation
the tidal force grows exponentially with a radial scale length
that is inversely proportional to the accretion rate.
The reason the singularity is weak and null in the ``standard'' picture
of an isolated black hole is that the accretion rate goes to zero,
so the growth rate of the tidal force becomes infinite.
The growth takes place over such a small proper time
that the tidal force becomes infinite before it has caused the
spacetime to deform significantly
\cite{Ori:2001pc}.
This is the weak singularity.
But in any real black hole,
the accretion rate, however tiny, is never zero.
For any finite accretion rate,
the outcome of the growing tidal force is deformation leading to collapse,
not a weak null singularity.

Even if the accretion rate were vanishingly small,
the diverging tidal force associated with the weak null singularity
would surely result in diverging pair creation,
and such pairs would surely act as an effective source of accretion,
again precipitating collapse.
\cite{Frolov:2006is}
have shown that in the simplified case of
a 1+1-dimensional charged black hole,
if the effects of pair creation of charged particles are taken into account,
then the result is collapse to a spacelike singularity
rather than a null singularity on the Cauchy horizon.
The effects of pair creation on mass inflation in black holes
in 4 spacetime dimensions has yet to be explored in the literature.

There is a complication to the
argument that the outgoing and ingoing streams that drive inflation
are dominated by accretion,
not by primeval Price tails.
Accreted particles must necessarily be ingoing at the outer horizon.
Thanks to centrifugal force,
freely falling particles that are sufficiently prograde
and sufficiently close to the equator
become outgoing at the inner horizon,
while other particles
remain ingoing at the inner horizon.
Thus at sufficiently low latitudes,
both outgoing and ingoing streams near the inner horizon can be fueled
by direct accretion of freely falling particles from outside the outer horizon.
However, for rotating black holes, there is a latitude above which
all particles that free fall from outside the outer horizon
remain ingoing at the inner horizon.
The critical latitude is highest for massless particles.
For massless particles that free fall from infinity,
the latitude above which all particles are ingoing at the inner horizon
ranges from $90^\circ$ (i.e.\ all latitudes are accessible)
for a non-rotating black hole,
to
$\sin^{-1} \! \sqrt{- 3 + 2 \sqrt{3}} \approx 43^\circ$
for a maximally-rotating (extremal) black hole.

Although there is a latitude above which
particles freely falling from outside the outer horizon cannot become outgoing,
in real astronomical black holes
collisional process such as electron-photon scattering
will inevitably result in outgoing particles accreting on to the inner horizon
at any latitude.
Even absent collisional processes,
accretion at different rates over the inner horizon
will perturb the geometry of the black hole away from a no-hair configuration.
The perturbation will result in gravitational waves that relax
the black hole back towards a no-hair configuration.
These gravitational waves will provide a source of both outgoing
and ingoing gravitational radiation at all latitudes on the inner horizon.
One way or another, directly or indirectly,
in any real astronomical black hole,
ongoing accretion will provide the dominant source of
outgoing and ingoing energy at all latitudes on the inner horizon.

The purpose of the present paper is to explore numerically
inflation and collapse in an accreting, rotating black hole,
in fully nonlinear general relativity.
Generically this is a challenging problem.
One of the challenges is that the numerical method
must be robust enough to follow the physical inflationary instability
without allowing unphysical gauge instabilities
to overwhelm the calculation.
Part of the intent of this paper is to test a new covariant
Hamiltonian tetrad formalism for general relativity coupled to matter fields,
developed by the author
\cite{Hamilton:2016mmc}
to meet the challenges.

To give confidence that the numerical results can be trusted,
it makes sense to start with the simplest computation
consistent with the goal of following
all 6 physical degrees of freedom of the gravitational field
through inflation and collapse.
The simplest computation is for a slowly accreting black hole,
where the geometry is almost Kerr down to near the inner horizon.

In the asymptotic limit of small accretion rates,
there exist conformally separable solutions
in which inflation occurs
\cite{Hamilton:2010a,Hamilton:2010b,Hamilton:2010c}.
The Kerr-Newman geometry has the property,
discovered by Carter \cite{Carter:1968c},
that the equations of motion of particles are Hamilton-Jacobi separable.
Conformal separability imposes the weaker condition that
the equations of motion are Hamilton-Jacobi separable
only for massless, not massive, particles.
The conformally separable solutions found by
\cite{Hamilton:2010a,Hamilton:2010b,Hamilton:2010c}
are axisymmetric and conformally time-translation invariant
(that is, self-similar).
Whereas strictly separable solutions (Kerr-Newman)
are time-translation invariant and therefore admit no accretion flow,
conformally separable solutions allow a conformally growing geometry,
and thus admit accretion, hence inflation.

The conformally separable solutions of
\cite{Hamilton:2010a,Hamilton:2010b,Hamilton:2010c}
require a special ``monopole'' accretion rate,
in which the outgoing and ingoing accretion flows on to the inner horizon
are uniform in latitude.
This might seem to limit the applicability of the solutions.
However,
because the mass inflation takes place over a proper time
that is short for small accretion rates,
and becomes shorter as inflation develops,
what happens at one point on the inner horizon becomes progressively
causally disconnected from what happens at other points.
During inflation,
outgoing and ingoing accretion flows,
regardless of their initial orbital parameters,
focus ever more narrowly
along the principal outgoing and ingoing null directions.
For small accretion rates,
the counterstreaming outgoing and ingoing beams
become hyper-relativistic while the geometry is still scarcely
perturbed from the Kerr geometry.
By the time the streams start to back-react on the geometry,
the transverse size of the causal patch of an inflating region is tiny.
Thus transverse gradients in the accretion flow on to the inner horizon
have a subdominant influence.
The argument that transverse gradients are subdominant
echoes a similar argument by BKL,
that spatial gradients become dominated by time gradients
during BKL collapse.

The conformally separable solutions predict that inflation is
followed by collapse,
but the solutions fail at some point during collapse
because of growing rotational motions.
A motivation for the present paper was to explore numerically what happens then.
As will be seen, the result is BKL collapse.
The inflationary and collapse phases
of the conformally separable solution
prove to be the first and second Kasner epochs of BKL collapse.

Units in this paper are geometric, $c = \kappa \, ( \equiv 8\pi G ) = 1$.

\section{BKL bounces and Kasner epochs}
\label{BKL-sec}

The 3+1 (ADM \cite{ADM:1959,ADM:1963}) formalism
shows that the 6 physical degrees of freedom of gravity
are encoded in the 6 components of the spatial metric,
a $3 \times 3$ symmetric matrix.
The spatial metric can be pictured as an ellipsoid,
characterized by the lengths of its three axes,
and by three rotation angles.
In BKL collapse,
the volume of the ellipsoid (determinant of the spatial metric)
decreases monotonically to zero in a finite time,
but one axis always expands while the other two collapse.
When one of the collapsing axes has collapsed to
sufficiently small size,
it ``bounces,'' turning from collapse into expansion,
while the previously expanding axis turns around and starts collapsing.
The directions of the three axes also change during a bounce.
The reason for BKL bounces is that
the Einstein equations (eq.~(\ref{Einsteineqaltformd}))
involve a potential energy term
proportional to squares of connection coefficients,
which provides a repulsive potential that diverges as any axis becomes small.
The sensitivity to initial conditions makes the behavior chaotic.

BKL refer to the epochs between BKL bounces as ``Kasner epochs''
since between bounces the spacetime evolves
in accordance with a vacuum solution of Einstein's equations
discovered by Kasner in 1921 \cite{Kasner:1921},
\begin{equation}
\label{kasnerlineelement}
  \dd s^2
  =
  - \,
  \dd T^2
  +
  \sum_{a = 1}^{3}
  ( a_a \dd X^a )^2
  \ ,
\end{equation}
in which the scale factors $a_a$ evolve as power laws with time $T$
\begin{equation}
  a_a \propto |T|^{q_a}
  \ ,
\end{equation}
with exponents $q_a$ satisfying
\begin{equation}
\label{kasnerexponents}
  \sum_{a = 1}^{3}
  q_a = 1
  \ , \quad
  \sum_{a = 1}^{3}
  ( q_a )^2 = 1
  \ .
\end{equation}
A parametric solution for the exponents $q_a$ is
\begin{equation}
   q_1
   =
   {- u \over 1 + u + u^2}
   \, , \ 
   q_2
   =
   {1 + u \over 1 + u + u^2}
   \, , \ 
   q_3
   =
   {u ( u + 1 ) \over 1 + u + u^2}
   \, .
\end{equation}
If the exponenents $q_a$ are ordered such that
$q_1 \leq q_2 \leq q_3$, then
\begin{equation}
\label{bianchiporder}
  - \tfrac{1}{3}
  \leq
  q_1
  \leq
  0
  \leq
  q_2
  \leq
  \tfrac{2}{3}
  \leq
  q_3
  \leq
  1
  \ .
\end{equation}

\section{Method}

\subsection{Numerical method}

The numerical method is described by \cite{Hamilton:2016mmc},
to which the reader is referred for details.
The degrees of freedom of the gravitational field are contained in
the line interval $\be$,
a vector 1-form with $4 \times 4 = 16$ degrees of freedom,
and
the Lorentz connection $\bGamma$,
a bivector 1-form with $6 \times 4 = 24$ degrees of freedom,
\begin{subequations}
\begin{align}
\label{eformdef}
  \be
  &\equiv
  e_{k{\kappa}}
  \, \bgamma^k
  \, \dd x^{\kappa}
  \ ,
\\
\label{Gammaformdef}
  \bGamma
  &\equiv
  \Gamma_{kl{\kappa}}
  \,
  \bgamma^k \wedgie \bgamma^l
  \,
  \dd x^{\kappa}
  \ .
\end{align}
\end{subequations}
The coefficients $e_{k\kappa}$ of the line interval
are commonly called the vierbein coefficients,
while the coefficients $\Gamma_{kl{\kappa}}$
of the Lorentz connection
are commonly called Ricci rotation coefficients,
or, especially when referred to a Newman-Penrose double-null tetrad,
spin coefficients.
The momentum canonically conjugate to the line interval $\be$ is
the 24-component pseudovector 2-form
$\bpi$
defined by
\begin{equation}
\label{bpidef}
  \bpi \equiv - \be \wedgie \bGamma
  \ .
\end{equation}
The momentum $\bpi$ is invertibly related to the Lorentz connection $\bGamma$.
The $16 + 24 = 40$ gravitational coordinates and momenta
$\be$ and $\bpi$
are governed by 40 Hamilton's equations
(eqs.~(19) of \cite{Hamilton:2016mmc}),
\begin{subequations}
\label{eqaltformd}
\begin{alignat}{2}
\label{torsioneqaltformd}
  \mbox{24 eqs:}
  &\quad&
  \bS
  \equiv
  \dext \be + \tfrac{1}{2} [ \bGamma , \be ]
  &=
  \kappa \tilde{\bSpin}
  \ ,
\\
\label{Einsteineqaltformd}
  \mbox{16 eqs:}
  &\quad&
  \bPi
  \equiv
  \dext \bpi + \tfrac{1}{2} [ \bGamma , \bpi ]
  -
  \tfrac{1}{4}
  \be \wedgie [ \bGamma , \bGamma ]
  &=
  \kappa \tilde{\bT}
  \ ,
\end{alignat}
\end{subequations}
where $\tilde{\bSpin}$ and $\tilde{\bT}$
are the (modified) spin angular-momentum and energy-momentum of matter.
Equations~(\ref{Einsteineqaltformd}) are the Einstein equations.
The $\tfrac{1}{4} \be \wedgie [ \bGamma , \bGamma ]$ term
in equation~(\ref{Einsteineqaltformd})
is the potential energy term that leads to BKL bounces in BKL collapse.


The gravitational coordinates $\be$ and momenta $\bpi$ contain excess degrees
of freedom, for two reasons.
First,
there are gauge degrees of freedom arising from the symmetries
of general relativity,
namely Lorentz transformations and coordinate transformations;
and second,
there are redundant degrees of freedom arising
from the 4-dimensional covariant description of the coordinates and momenta.

The standard way to remove the excess degrees of freedom
from the equations of motion
is to perform a 3+1 space+time split of spacetime.
The split decomposes the 16-component line interval
into
4 time components $\be_{\tform}$
(subscripted $\tform$),
and
12 spatial components $\be_{\alphaform}$
(subscripted $\alphaform$),
and the conjugate momentum
into
12 time components $\bpi_{\tform}$
and
12 spatial components $\bpi_{\alphaform}$,
\begin{equation}
  \be = \be_{\tform} + \be_{\alphaform}
  \ , \quad
  \bpi = \bpi_{\tform} + \bpi_{\alphaform}
  \ .
\end{equation}
The $\tform$ and $\alphaform$
subscripts should be interpreted as labels, not indices.
After the 3+1 space+time split,
the physical degrees of freedom of the gravitational field
comprise the 12 spatial components $\be_{\alphaform}$ of the line interval
and the 12 spatial components $\bpi_{\alphaform}$ of their conjugate momenta.
The 4 time components $\be_{\tform}$ of the line interval
(the lapse and shift)
are gauge degrees of freedom arising from symmetry under
coordinate transformations,
while the 12 time components $\bpi_{\tform}$ of the momentum
comprise 6 gauge degrees of freedom associated with symmetry
under Lorentz transformations,
and another 6 degrees of freedom that
are simply redundant
(see \S{}II\,D.E of
\cite{Hamilton:2016mmc}).

Thus Hamilton's equations~(\ref{eqaltformd})
split into $12 + 12= 24$ equations of motion involving time derivatives,
together with
6 Gaussian constraints,
6 identities,
and 4 Hamiltonian constraints,
\begin{subequations}
\label{eqaltformtRS}
\begin{alignat}{2}
\label{torsionealtformtS}
  \mbox{12 eqs of mot:}
  &\quad&
  \bS_{\tform}
  &=
  \kappa \tilde{\bSpin}_{\tform}
  \ ,
\\
\label{EinsteineqaltformtR}
  \mbox{12 eqs of mot:}
  &\quad&
  \bPi_{\tform}
  &=
  \kappa \tilde{\bT}_{\tform}
  \ ,
\\
\label{torsionconstraintfaltformtS}
  \mbox{6 Gauss + 6 ids:}
  &\quad&
  \bS_{\alphaform}
  &=
  \kappa \tilde{\bSpin}_{\alphaform}
  \ ,
\\
\label{EinsteinconstraintfaltformtR}
  \mbox{4 Ham:}
  &\quad&
  \bPi_{\alphaform}
  &=
  \kappa \tilde{\bT}_{\alphaform}
  \ .
\end{alignat}
\end{subequations}
Equations~(\ref{torsionealtformtS}) and~(\ref{EinsteineqaltformtR})
are equations of motion in the sense
that they determine the time derivatives
$\dext_{\tform} \be_{\alphaform}$
and
$\dext_{\tform} \bpi_{\alphaform}$
of the gravitational coordinates
$\be_{\alphaform}$ and momenta $\bpi_{\alphaform}$.
The time derivative here is the 1-form
$\dext_{\tform} \equiv ( \partial / \partial T ) \dd T$
where $T$ is a suitable time coordinate.
\cite{Hamilton:2016mmc}
shows how to isolate the 6 Gaussian constraints and 6 identities
in equations~(\ref{torsionconstraintfaltformtS}).
The 6 identities define a 6-component traceless spatial tensor that
\cite{Hamilton:2016mmc}
calls the ``gravitational magnetic field''
since its role in the gravitational equations is similar to
that of the magnetic field in the equations of electromagnetism.


\subsection{Conformally separable initial conditions}

The initial conditions adopted in this paper are those
of the conformally separable solution
\cite{Hamilton:2010a,Hamilton:2010b}
for a rotating, slowly accreting black hole.
The conformally separable solution is not exact,
but rather holds in the vicinity of the inner horizon
in the asymptotic limit of small accretion rates.

The conformally separable solution is sourced by two collisionless
null streams, one outgoing and one ingoing.
The solution is parameterized by three dimensionless parameters,
which are the spin parameter $a$ of the rotating black hole,
and two constants $u$ and $v$.
The accretion rates of the outgoing and ingoing streams
are proportional to $u \pm v$
(see eq.~\ref{Npm}).
Physically realistic solutions have positive accretion rates;
small positive accretion rates require that the constants $v$ and $u$ satisfy
\begin{equation}
  0 < v < u \ll 1
  \ .
\end{equation}

The advantage of
choosing the initial conditions
to be those of the conformally separable solution
is that it is possible to start
the integration inward into the black hole near the inner horizon,
at a stage of evolution where inflation is already well advanced.
The behavior of the accretion flow
simplifies greatly near the inner horizon,
because outgoing and ingoing accretion streams focus along
just two directions, the outgoing and ingoing principal null directions,
regardless of their initial orbital parameters.

Choosing the initial conditions
to be those of the conformally separable solution
has the additional merit that agreement between
the solution and numerical calculations
gives confidence that both are reliable.

With respect to Boyer-Lindquist coordinates
$x^\mu \equiv \{ r , t , \theta , \phi \}$,
the conformally separable Kerr line element
may be written
\begin{align}
\label{knlineelement}
  \dd s^2
  =
  \rhosep^2 \ee^{2 v t}
  \Biggl\{
  \left[
  {\ee^{- 5\xi} \dd r^2 \over R^4 \Delta_r}
  -
  {\ee^{\xi} \Delta_r \over ( 1 - \omega_r \omega_\theta )^2}
  \left( \dd t - \omega_\theta \dd \phi \right)^2
  \right]
  &
\nonumber
\\
  \mbox{}
  +
  \ee^{-2\xi}
  \left[
  \dd \theta^2
  +
  {\Delta_\theta \over ( 1 - \omega_r \omega_\theta )^2}
  \left( \dd \phi - \omega_r \dd t \right)^2
  \right]
  \Biggr\}
  \ ,
  &
\end{align}
where $\rhosep$ is the separable conformal factor
\begin{equation}
\label{rhosep}
  \rhosep
  =
  \sqrt{r^2 + a^2 \cos^2\!\theta}
  \ ,
\end{equation}
$\omega_r$ and $\omega_\theta$ are functions respectively only of
$r$ and $\theta$,
\begin{equation}
  \omega_r =
  {a \over R^2}
  \ , \quad
  R \equiv \sqrt{r^2 + a^2}
  \ , \quad
  \omega_\theta
  =
  a \sin^2\!\theta
  \ ,
\end{equation}
and the horizon function
$\Delta_r$
and polar function
$\Delta_\theta$
are similarly functions respectively only of
$r$ and $\theta$.
In the conformally separable solution,
the polar function $\Delta_\theta$ always takes its Kerr value,
but the horizon function $\Delta_r$,
which satisfies the equation of motion~(\ref{dDelta}),
equals its Kerr value $\mathring{\Delta}_r$
only at radii sufficiently above the inner horizon,
$r \gg r_-$,
\begin{equation}
\label{Deltarth}
  \Delta_r \overset{r \gg r_-}{\longrightarrow}
  \mathring{\Delta}_r =
  {1 \over R^2} \left( 1 - {2 M r \over R^2} \right)
  \ , \quad
  \Delta_\theta
  =
  \sin^2\!\theta
  \ .
\end{equation}
Inside the outer horizon,
the horizon function $\Delta_r$ is negative,
the radial coordinate $r$ is timelike,
and the time coordinate $t$ is spacelike.
The horizon function in
\cite{Hamilton:2010a,Hamilton:2010b},
written there as $\Delta_x$,
is $\Delta_x = \ee^{3\xi} \Delta_r$
where $\xi$ is the inflationary exponent given by equations~(\ref{xiU}).
The convention for the horizon function adopted by
\cite{Hamilton:2010a,Hamilton:2010b}
is more natural from the perspective of separating the Einstein equations,
but the present definition has the advantage that the horizon function
$\Delta_r$ varies slowly (``freezes'') in the collapse regime where
the inflationary exponent $\xi$ grows large.

The geometry described by
the line element~(\ref{knlineelement})
is conformally time-translation symmetric
(i.e.\ self-similar),
expanding as $\ee^{v t}$.
The radial coordinate $r$ is a conformal (self-similar) coordinate,
and likewise $M$ is the conformal mass of the black hole.
The proper mass and radius of the black hole increase
as $\ee^{v t}$
as seen by a distant observer.

The conformally separable Boyer-Lindquist line element~(\ref{knlineelement})
defines not only a metric, but also, through
\begin{equation}
\label{lineelement}
  \dd s^2
  =
  e^k{}_\mu e^l{}_\nu
  \,
  \bgamma_k \cdot \bgamma_l
  \,
  \dd x^\mu \dd x^\nu
  \ ,
\end{equation}
a vierbein $e^k{}_\mu$, and an associated locally inertial tetrad
$\bgamma_k \equiv \{ \bgamma_r , \bgamma_t , \bgamma_\theta , \bgamma_\phi \}$,
whose scalar products form the Minkowski metric,
$\bgamma_k \cdot \bgamma_l = \eta_{kl}$.
Corresponding to the orthonormal tetrad is
a double-null Newman-Penrose tetrad
$\{ \bgamma_v , \bgamma_u , \bgamma_+ , \bgamma_- \}$
defined by
\begin{equation}
  \bgamma_{\overset{\scriptstyle v}{\scriptstyle u}}
  \equiv
  \frac{1}{\sqrt{2}}
  ( \bgamma_r \pm \bgamma_t )
  \ , \quad
  \bgamma_\pm
  \equiv
  \frac{1}{\sqrt{2}}
  ( \bgamma_\theta \pm \im \bgamma_\phi )
  \ .
\end{equation}
The nonvanishing scalar products of the Newman-Penrose tetrad are
$\bgamma_v \cdot \bgamma_u = -1$ and
$\bgamma_+ \cdot \bgamma_- = 1$.
The Boyer-Lindquist tetrad is constructed precisely so that
the null directions $\bgamma_v$ and $\bgamma_u$
point along respectively the principle outgoing and ingoing null directions
of the black hole.

Outgoing and ingoing principal null geodesics follow
$\theta = \mbox{constant}$, $\dd \phi / \dd t = \omega_r$,
and $r^\ast {\pm}\, t = \mbox{constant}$,
where $r^\ast$ is the tortoise coordinate.
The tortoise coordinate $r^\ast$ satisfies
\begin{equation}
  \dd r^\ast
  =
  {\ee^{- 3 \xi } \dd r \over R^2 \Delta_r}
  \ ,
\end{equation}
which is a function only of conformal radius $r$.
The tortoise coordinate $r^\ast$
increases inward towards the inner horizon.

Separation of Einstein's equations
\cite{Hamilton:2010a,Hamilton:2010b}
sourced by outgoing and ingoing collisionless null streams
leads to the following equations governing the evolution of
the inflationary exponent $\xi$
and the horizon function $\Delta_r$
in the line element~(\ref{knlineelement}),
\begin{subequations}
\label{dxiU}
\begin{align}
\label{dxi}
  {\dd \xi \over \dd r^\ast}
  &\equiv
  U
  \ ,
\\
\label{dU}
  {\dd U \over \dd r^\ast}
  &=
  2 ( U^2 - \vel^2 )
  \ ,
\\
\label{dDelta}
  {\dd \ln | R^4 \Delta_r / r | \over \dd r^\ast}
  &=
  {r^2 - a^2 \over r R^2}
  \ ,
\end{align}
\end{subequations}
with initial conditions
$\xi = 0$, $U = u$, and
$\Delta_r = \mathring{\Delta}_r$.
Equations~(\ref{dxi}) and~(\ref{dU}) solve to
\begin{subequations}
\label{xiU}
\begin{align}
\label{xi}
  &\xi
  =
  \frac{1}{4}
  \ln \left( {U^2 - \vel^2 \over \uel^2 - \vel^2} \right)
  \ ,
\\
\label{U}
  &{( U - \vel ) ( \uel + \vel ) \over ( U + \vel ) ( \uel - \vel )}
  =
  \ee^{4 v r^\ast}
  \ .
\end{align}
\end{subequations}
\cite{Hamilton:2010a,Hamilton:2010b}
give a solution of equation~(\ref{dDelta}) for the horizon function
$\Delta_r$
valid in the approximation that the radius $r$ is frozen
at its inner horizon value $r_-$,
which is valid in the asymptotic limit of small accretion rates.
In the present paper, where accretion rates are finite,
the horizon function $\Delta_r$
is solved instead from its differential equation~(\ref{dDelta})
with the Kerr initial condition~(\ref{Deltarth}),
so that the integration can be started from a (small but) finite distance
above the inner horizon,
and consistency between integrations from different starting radii can be tested.


\subsection{Energy-momenta}

The accretion flow consists of two
null, pressureless, collisionless streams,
one outgoing ($+$) and one ingoing ($-$).
The tetrad-frame energy-momentum tensor $T_{kl}$
is a sum over the two collisionless streams
(see \S{}VII of \cite{Hamilton:2010b}),
\begin{equation}
\label{Tkl}
  T_{kl}
  =
  n^+_k p^+_l
  +
  n^-_k p^-_l
  \ ,
\end{equation}
where $p^\pm_k$
and $n^\pm_k$ are the outgoing and ingoing momenta and number currents
\begin{equation}
\label{nk}
  n^\pm_k
  =
  N^\pm
  p^\pm_k
  \ ,
\end{equation}
with $N^\pm$ being outgoing and ingoing scalar densities.
The equations of motion
for each of the two streams
are the geodesic equation
\begin{equation}
\label{dp}
  p^\pm_k D^k p^\pm_l = 0
  \ ,
\end{equation}
and number conservation
\begin{equation}
\label{dn}
  D^k n^\pm_k = 0
  \ .
\end{equation}

In the conformally separable initial conditions,
the tetrad-frame outgoing and ingoing momenta $p^{\pm}_k$ are
related to Hamilton-Jacobi parameters $P^\pm_k$ by
\begin{equation}
  p^\pm_1
  =
  {P^\pm_1 \over \rhosep \sqrt{\Delta_r} \ee^{\xi/2}}
  \ , \quad
  p^\pm_k
  =
  {P^\pm_k \over \rhosep \sqrt{\Delta_\theta} \ee^{-\xi}}
  \ \ 
  ( k = 2, 3 )
  \ .
\end{equation}
The Hamilton-Jacobi parameters $P^\pm_k$ of the
outgoing and ingoing collisionless streams are,
equation~(22) of \cite{Hamilton:2010a},
\begin{subequations}
\label{HamJacP}
\begin{align}
\label{HamJacP1}
  P^\pm_1
  &=
  \pm \Delta_r^\prime - \vel
  \ ,
\\
\label{HamJacP2}
  P^\pm_2
  &=
  {2 \Delta_\theta \over \sin\theta} {\partial \rhosep \over \partial \theta}
  \ ,
\\
\label{HamJacP3}
  P^\pm_3
  &=
  \mp
  {R^2 \Delta_\theta \over 1 - \omega_r \omega_\theta}
  {\partial \omega_r \over \partial r}
  +
  2 \vel \omega_\theta
  \ ,
\end{align}
\end{subequations}
where $\Delta_r^\prime \equiv - R^2 \dd \Delta_r / \dd r$.
The outgoing and ingoing densities $N^\pm$
in the conformally separable initial conditions are
\begin{equation}
\label{Npm}
  N^\pm
  =
  {U \pm \vel \over 2 ( \Delta_r^\prime \mp \vel )}
  \ .
\end{equation}

In practice, because the conformally separable solution is not exact,
the densities and momenta of the collisionless streams in the initial conditions
are adjusted slightly so that the energy-momentum tensor
agrees as well as possible with the Einstein tensor deduced
from the line element~(\ref{knlineelement}).

Initially the outgoing and ingoing momenta are closely
(though not exactly) aligned with
the outgoing and ingoing principal null directions;
the momenta become misaligned with the principal directions at later times.

The conformally separable solution assumes zero spin-angular momentum
$\tilde{\bSpin}$,
hence vanishing torsion $\bS$,
as is the common assumption in general relativity.

\subsection{Integrate numerically along a single radial direction}
\label{singledirection-sec}

For simplicity,
this paper adopts the approach of integrating inward into the black hole
along a single radial direction, at a definite latitude.
Gradients in the spatial directions $t, \theta, \phi$
are taken to be given by the conformally separable solution.
The assumption of axisymmetry means that gradients in the azimuthal direction
$\phi$ vanish identically,
while the assumption of conformal time symmetry
means that derivatives
$\partial \ln e_{k\mu} / \partial t = \vel$
of logarithmic vierbein elements with respect to $t$
are all equal to the constant $\vel$.
Gradients in the latitude direction $\theta$ are nontrivial.

The approximation that angular gradients are those of the conformally
separable solution is at least partially tested
by the degree to which the Hamiltonian
and Gaussian constraints are satisfied.

\subsection{Factor vierbein into dynamical and fixed parts}
\label{factorvierbein-sec}

The vierbein $e_{k\mu}$
defined by the conformally separable line element~(\ref{lineelement})
can be written as the product of a dynamical vierbein
$\tilde{e}_{k\kappa}$
that is a function only of the timelike coordinate $r$,
and a fixed vierbein
$\mathring{e}^\kappa{}_\mu$
whose elements are fixed to those of the parent Kerr black hole,
\begin{equation}
\label{doublevierbein}
  e_{k\mu}
  =
  \tilde{e}_{k\kappa}
  \mathring{e}^\kappa{}_\mu
  \ .
\end{equation}
The fixed vierbein
$\mathring{e}^\kappa{}_\mu$
with respect to Boyer-Lindquist coordinates
$x^\mu \equiv \{ r , t , \theta , \phi \}$ is
\begin{equation}
\label{kerrvierbein}
  \mathring{e}^\kappa{}_\mu
  \equiv
  \rhosep
  \ee^{v t}
  \left(
  \begin{array}{cccc}
  1 & 0 & 0 & 0 \\
  0 & \displaystyle {1 \over 1 - \omega_r \omega_\theta} & 0 & \displaystyle {- \omega_\theta \over 1 - \omega_r \omega_\theta} \\
  0 & 0 & 1 & 0 \\
  0 & \displaystyle {- \omega_r \sqrt{\Delta_\theta} \over 1 - \omega_r \omega_\theta} & 0 & \displaystyle {\sqrt{\Delta_\theta} \over 1 - \omega_r \omega_\theta}
  \end{array}
  \right)
  \ ,
\end{equation}
which serves to align the tetrad with the principle null directions
of the black hole.
In the conformally separable initial conditions,
the dynamical vierbein is the diagonal matrix
\begin{equation}
\label{dynvierbein}
  \tilde{e}_{k\kappa}
  \equiv
  \left(
  \begin{array}{cccc}
  \displaystyle {\ee^{- 5 \xi / 2} \over R^2 \sqrt{| \Delta_r |}} & 0 & 0 & 0 \\
  0 & \displaystyle \ee^{\xi / 2} \sqrt{| \Delta_r |} & 0 & 0 \\
  0 & 0 & \displaystyle \ee^{- \xi} & 0 \\
  0 & 0 & 0 & \displaystyle \ee^{- \xi}
  \end{array}
  \right)
  \ .
\end{equation}
As the numerical integration proceeds inward,
the dynamical vierbein $\tilde{e}_{k\kappa}$ ceases to be diagonal.

The numerical integration works with a time variable $T$
related to the timelike radial coordinate $r$ by
\begin{equation}
\label{dTdr}
  \alpha \, \dd T
  =
  -
  {\ee^{- 5 \xi / 2} \, \dd r \over R^2 \sqrt{| \Delta_r |}}
  =
  \ee^{\xi / 2} \sqrt{| \Delta_r |} \, \dd r^\ast
  \ ,
\end{equation}
where $\alpha$ is the dynamical lapse, a function only of $r$.
With respect to coordinates
$x^\mu \equiv \{ T , t , \theta , \phi \}$,
the conformally separable dynamical line interval~(\ref{dynvierbein}) is
\begin{equation}
\label{dynvierbeina}
  \tilde{e}_{k\kappa}
  \equiv
  \left(
  \begin{array}{cccc}
  - \alpha & 0 & 0 & 0 \\
  0 & a_1 & 0 & 0 \\
  0 & 0 & a_2 & 0 \\
  0 & 0 & 0 & a_3
  \end{array}
  \right)
  \ ,
\end{equation}
where the spatial scale factors $a_a$ are
\begin{equation}
  a_1 = \ee^{\xi / 2} \sqrt{| \Delta_r |}
  \ , \quad
  a_2 = a_3 = \ee^{- \xi}
  \ .
\end{equation}

By assumption,
the only part of the line element that depends on spatial variables
$t, \theta, \phi$
is the fixed part
$\mathring{e}^\kappa{}_\mu$,
equation~(\ref{kerrvierbein}).
The spatial variation,
which is assumed to continue to be given by equation~(\ref{kerrvierbein})
even after the conformally separable solution breaks down,
in turn depends on the radius $r$.
The radius $r$ is taken to be given by a plausible extrapolation of
equation~(\ref{dTdr}),
\begin{equation}
\label{dr}
  {\dd r \over \dd T}
  =
  -
  {\alpha R^2 a_1 \over a_2 a_3}
  \ ,
\end{equation}
with the scale factor $a_a$ in each tetrad direction $a = 1, 2, 3$
being taken to be determined by the expansion $\Gamma^a_{0a}$
(no sum over $a$)
in that direction $a$,
\begin{equation}
\label{daa}
  {\dd a_a \over \dd T}
  =
  \alpha
  R^2
  \Gamma^a_{0a}
  \quad
  \mbox{no sum over $a$}
  \ .
\end{equation}
Equations~(\ref{dr}) and~(\ref{daa})
are not part of the system of equations~(\ref{eqaltformd})
governing the evolution of the gravitational field.
Rather,
equations~(\ref{dr}) and~(\ref{daa}) are guesses
necessitated by the simplifying approximation adopted in this paper
that gradients in angular directions may be approximated adequately
by (a plausible extrapolation of) the conformally separable solution.
The justification for this approximation is the BKL argument that
transverse gradients are subdominant to gradients in the time direction
during BKL collapse.
The validity of the approximation can be tested, at least in part,
by the extent to which the (Hamiltonian and Gaussian) constraints
are satisfied (see \S\ref{Ham-sec}).

\subsection{First two Kasner epochs}

The conformally separable dynamical vierbein~(\ref{dynvierbeina})
has the form of two successive Kasner epochs.
As discussed by
\cite{Hamilton:2010a},
the traditional inflationary regime introduced by Poisson \& Israel
\cite{Poisson:1990eh}
corresponds to the situation where the first and second radial gradients
$\partial \xi / \partial r$
and
$\partial^2 \xi / \partial r^2$
of the inflationary exponent grow large,
while the inflationary exponent $\xi$ itself scarcely budges from zero.
In this inflationary regime where
the inflationary exponent remains close to zero,
$\xi \approx 0$,
the dynamical vierbein~(\ref{dynvierbeina}) near the inner horizon
looks like a Kasner spacetime~(\ref{kasnerlineelement}) with exponents
\begin{equation}
\label{kasnerqepoch1}
  q_1 = 1
  \ , \quad
  q_2 = q_3 = 0
  \ ,
\end{equation}
meaning that the spacetime is collapsing along the radial direction only.
For a black hole that continues to accrete,
as is always true in an astronomically realistic black hole,
the focusing of accretion streams along the principal outgoing and ingoing
null directions produces a tidal force that eventually causes
inflation to stall,
and the geometry to start collapsing along the transverse directions
and expanding along the radial direction.
This corresponds to the collapse phase of the conformally separable solution,
where the inflationary exponent $\xi$ grows large
while the horizon function $\Delta_r$ freezes to a constant.
In this collapse regime,
the dynamical vierbein~(\ref{dynvierbeina})
looks like a second Kasner spacetime~(\ref{kasnerlineelement})
with exponents
\begin{equation}
\label{kasnerqepoch2}
  q_1 = - \tfrac{1}{3}
  \ , \quad
  q_2 = q_3 = \tfrac{2}{2}
  \ .
\end{equation}
Thus the conformally separable solution appears to describe
the first two Kasner epochs of BKL-like collapse.

The Schwarzschild solution for a non-rotating black hole has no inner horizon.
Near its singular surface (the singularity is a spacelike surface, not a point),
the Schwarzschild solution resembles
a Kasner spacetime with exponents~(\ref{kasnerqepoch2}),
coinciding with those of a rotating black hole in the second Kasner epoch.

\subsection{Aligning the Lorentz frame with the principal null frame}

The tetrad defined by
the conformally separable line element~(\ref{knlineelement})
is aligned with the principle null directions.
An observer at rest in the tetrad frame
sees outgoing and ingoing principal null streams
to be $180^\circ$ apart on the sky
(outgoing appears below, ingoing above),
equally blueshifted,
and equally rotated.
The condition that the principle null directions be geodesic
(their transverse momenta vanish, $p_2 = p_3 = 0$)
imposes the following 4 conditions on the tetrad-frame Lorentz connections
$\Gamma_{klm}$
(related to the components of the bivector 1-form Lorentz connection by
$\Gamma_{klm} = e_m{}^\mu \Gamma_{kl\mu}$),
\begin{subequations}
\label{gauge0a}
\begin{align}
\label{gauge0}
  \Gamma_{020} + \Gamma_{121}
  =
  \Gamma_{030} + \Gamma_{131}
  &=
  0
  \ ,
\\
\label{gaugea}
  \Gamma_{021} + \Gamma_{120}
  =
  \Gamma_{031} + \Gamma_{130}
  &=
  0
  \ .
\end{align}
\end{subequations}
The condition that outgoing and ingoing streams
along the principle null directions
appear equally blueshifted (their energies $p^0$ are the same) is
\begin{equation}
\label{gaugeblue}
  \Gamma_{010}
  +
  ( e^{1T} / e^{0T} ) \Gamma_{110}
  =
  - v e^{1t}
  \ .
\end{equation}
Equation~(\ref{gaugeblue}) holds for any
spacetime that is conformally time-translation invariant.
The condition that the outgoing and ingoing streams
along the principle null directions
appear equally rotated about the null directions is
\begin{equation}
\label{gaugerot}
  \Gamma_{230}
  =
  0
  \ .
\end{equation}

In the conformally separable solution,
the principal null directions are geodesic.
It is natural to define a ``principal frame''
by the geodesic continuation of the principal null directions.

\subsection{Gauge choices}
\label{gauge-sec}

There are 10 gauge choices to be made,
4 associated with symmetry under coordinate transformations,
and 6 associated with symmetry under Lorentz transformations.

The 4 components of the time $T$ component
$\tilde{e}_{kT}$
of the dynamical vierbein,
commonly called the (dynamical) lapse $\alpha$ and shift $\beta_a$,
can be treated as gauge variables
arbitrarily adjustable under a coordinate transformation.
This paper chooses
the dynamical lapse to be 1 and the dynamical shift to be zero,
\begin{equation}
\label{lapseshift}
  \alpha \equiv \tilde{e}_{0T} = 1
  \ , \quad
  \beta_a \equiv \tilde{e}_{aT} = 0 \quad (a = 1,2,3)
  \ .
\end{equation}


For the remaining 6 gauge choices,
this paper tries 2 different sets of gauge choices,
referred to here as
principal gauge,
and
BSSN
(Baumgarte-Shapiro-Shibata-Nakamura)
\cite{Baumgarte:2010,Brown:2012me}
gauge.
BSSN gauge
should perhaps be called BSSN-like,
because the numerical method followed in this paper
is still the covariant Hamiltonian tetrad approach of
\cite{Hamilton:2016mmc},
as opposed to the coordinate-based,
Lorentz-gauge-invariant approach commonly referred to as BSSN.
The key distinction between principal and BSSN gauges
is that whereas principal gauge imposes 6 conditions on
the Lorentz connections,
BSSN gauge replaces 3 of the 6 gauge conditions
by the 3 ADM gauge conditions~(\ref{admgauge}) on the vierbein.

\subsubsection{Principal gauge}
\label{princ-sec}

Principal gauge imposes the 6 conditions~(\ref{gauge0a})--(\ref{gaugerot})
on the Lorentz connections.
The conditions ensure that, in rest frame of the tetrad,
the (geodesic continuation of the)
outgoing and ingoing principal null streams
remain $180^\circ$ apart on the sky,
and equally blueshifted, and equally rotated.

\subsubsection{BSSN gauge}
\label{bssn-sec}

BSSN
\cite{Baumgarte:2010,Brown:2012me}
gauge
imposes the 3 ADM gauge conditions
\begin{equation}
\label{admgauge}
  e_{0\alpha} = 0 \quad (\alpha = t,\theta,\phi)
  \ ,
\end{equation}
on the vierbein.
The equations of motion for the 3 components
$e_{0\alpha}$
are then reinterpreted as identities,
equation~(45) of \cite{Hamilton:2016mmc}.
The ADM conditions~(\ref{admgauge})
use up 3 of the 6 Lorentz gauge freedoms on the vierbein.
For the remaining 3 Lorentz gauge freedoms, BSSN gauge in
this paper imposes the conditions~(\ref{gaugea}) and~(\ref{gaugerot}).

%

\subsection{Integration scheme}

The integrator is a 4th-order predictor-corrector (Adams-Bashforth).
The predictor-corrector routine calls two subroutines in succession,
a subroutine that
sets time derivatives of the 12 spatial coordinates $\be$
and 12 spatial momenta $\bpi$
in accordance with the equations of
motion~(\ref{torsionealtformtS}) and~(\ref{EinsteineqaltformtR}),
and a subroutine that applies gauge choices and identities to determine
the remaining coordinates and momenta,
namely the time components $\be_{\tform}$ and $\bpi_{\tform}$.

The routine that sets time derivatives has an option
to compute the evolution of the conformal parts of the line interval $\be$
and its conjugate momentum $\bpi$
separately from the nonconformal parts.
The nonconformal (dimensionless) parts of the spatial line interval
and its conjugate momentum are $\be / | e|^{1/3}$ and $\bpi / | e |^{1/3}$,
where $| e |$ is the determinant of the spatial vierbein.
The equations of motion for the conformal parts of $\be$ and $\bpi$
are equations~(33) and~(34) of \cite{Hamilton:2016mmc}
for the spatial volume element $\be^3$
and the spatial expansion $\bvartheta \equiv \tfrac{1}{2} \be \wedgie \bpi$.
The equation for the expansion $\bvartheta$ has an option to adjust it
by adding an arbitrary factor
of $\be_{\tform}$ wedged with the Hamiltonian constraints,
equation~(35) of \cite{Hamilton:2016mmc}.
In practice,
adjusting the equation of motion for
the expansion proves to have no appreciable effect on the results.
By default, therefore, the results reported in this paper
do not treat the conformal parts in any special way:
the line interval and conjugate momentum are left unscaled,
and the evolution of the conformal parts of $\be$ and $\bpi$ is computed from
equations~(\ref{torsionealtformtS}) and~(\ref{EinsteineqaltformtR})
just like the nonconformal parts.

After calling the subroutine that sets time derivatives of
the spatial coordinates and momenta $\be$ and $\bpi$,
the predictor-corrector routine calls a subroutine that sets
the 4 time components $\be_{\tform}$
and 12 time components $\bpi_{\tform}$.

The 4 time components $\be_{\tform}$,
the lapse and shift, are gauge choices, equations~(\ref{lapseshift}).

The
12 time components $\bpi_{\tform}$
and
12 spatial components $\bpi$
of the conjugate momentum together satisfy 30 equations,
including the 12 equations of motion that determine $\bpi$.
The 30 equations may expressed as the matrix equation
\begin{equation}
\label{ApiC}
  \left(
  \begin{array}{cc}
  A_{\tform} & A \\
  0 & 1
  \end{array}
  \right)
  \left(
  \begin{array}{c}
  \bpi_{\tform} \\ \bpi
  \end{array}
  \right)
  =
  \left(
  \begin{array}{cc}
  C \\ \bpi
  \end{array}
  \right)
  \ ,
\end{equation}
where $C$ is a vector with 18 entries, consisting of
\begin{equation}
  C
  =
  \left(
  \begin{array}{l}
  \mbox{6 (\mbox{Principal}) or 3 (\mbox{BSSN}) gauge choices}
  \\
  \mbox{6 Gaussian constraints}
  \\
  \mbox{6 (\mbox{Principal}) or 9 (\mbox{BSSN}) identities}
  \end{array}
  \right)
  \ .
\end{equation}
Given $\bpi$ from their equations of motion,
equation~(\ref{ApiC}) reduces to
\begin{equation}
\label{ApiTC}
  A_{\tform}
  \bpi_{\tform}
  =
  C
  -
  A \bpi
  \ .
\end{equation}
Equation~(\ref{ApiTC}) is a set of 18 equations
for the 12 time components $\bpi_{\tform}$ of the conjugate momentum.
The 18 equations include 6 redundant equations, the 6 Gaussian constraints.
The 6 redundant equations in equations~(\ref{ApiTC})
could potentially be eliminated by dropping the constraint equations.
That this may not be the best strategy
is evidenced by the fact that the constraint equation
associated with the ADM gauge condition~(\ref{admgauge})
is formally the BSSN momentum equation,
equation~(43) of \cite{Hamilton:2016mmc},
and keeping the BSSN momentum equation is precisely what,
apparently,
makes the BSSN approach superior to the traditional ADM approach.

Rather than attempt to determine which constraints to drop a priori,
this paper adopts the numerical strategy of solving
equation~(\ref{ApiTC})
for $\bpi_{\tform}$ by singular value decomposition (SVD).

\subsection{Numerical precision}

The results shown in this paper are shown up to the point that
the numerical integration grinds to a halt.
The numerical integration stalls
when the matrix equation~(\ref{ApiTC}) solved by singular value decomposition
becomes ill-conditioned, that is,
the ratio of the largest to smallest singular values
exceeds the numerical precision.

To mitigate loss of precision,
the code, written in c, is implemented in long double.
On x86 architecture,
a long double uses 80 bits,
comprising a sign,
64 bits in the mantissa, and 15 bits in the exponent,
whereas
an ordinary double uses 64 bits,
comprising a sign,
52 bits in the mantissa, and 11 bits in the exponent.
The numerical integration stalls when the condition number of
the singular-value matrix
$A_{\tform}$
exceeds the precision
$\sim 2^{64} \approx 10^{19}$.

\subsection{Decomposition of the line interval}

Only 6 of the 16 components
$e_{k\kappa}$
of the vierbein
represent physical, coordinate and Lorentz gauge-invariant
degrees of freedom of the gravitational field.
The 4 time components
$e_{k T}$
of the vierbein, the lapse and shift, are gauge variables,
adjustable by a coordinate transformation.
The remaining 12 components of the vierbein
are the spatial components
$e_{k\alpha}$.
Of these, the physical,
Lorentz gauge-invariant degrees of freedom
are the 6 components of the symmetric spatial metric
$g_{\alpha\beta} = e_{k\alpha} \eta^{kl} e_{l\beta}$.

Because of the block diagonal form (in time and space) of the fixed vierbein
$\mathring{e}^\kappa{}_\mu$
in the decomposition~(\ref{doublevierbein}) of the vierbein
into dynamical and fixed factors,
the degrees of freedom of the spatial vierbein
$e_{k\alpha}$
are the same as those of the spatial dynamical vierbein
$\tilde{e}_{k\alpha}$,
\begin{equation}
  e_{k\alpha}
  =
  \tilde{e}_{k\beta} \mathring{e}^\beta{}_\alpha
  \ , \quad
  \tilde{e}_{k\beta}
  =
  e_{k\alpha} \mathring{e}_\beta{}^\alpha
  \ .
\end{equation}
The physical, Lorentz gauge-invariant degrees of freedom
of the dynamical spatial vierbein
are those of the dynamical spatial metric
$\tilde{g}_{\alpha\beta} = \tilde{e}_{k\alpha} \eta^{kl} \tilde{e}_{l\beta}$.
In abbreviated notation
\begin{equation}
  \tilde{g} = \tilde{e}^\transpose \! \eta \tilde{e}
  \ .
\end{equation}
The $3 \times 3$ dynamical spatial metric $\tilde{g}$ can be decomposed as
\begin{equation}
\label{gUU}
  \tilde{g}
  =
  U^\transpose \! a a U
  \ ,
\end{equation}
where $a$ is a $3 \times 3$ diagonal matrix with all positive entries,
and $U$ is a $3 \times 3$ matrix that can be chosen in a variety of ways:
$U$ could be upper triangular with unit diagonals,
or lower triangular with unit diagonals,
or orthogonal.
The $4 \times 3$ spatial dynamical vierbein $\tilde{e}$ can then be written
\begin{equation}
\label{eLaU}
  \tilde{e}
  =
  L
  \left(
  \begin{array}{c}
  0 \\
  a
  \end{array}
  \right)
  U
  \ ,
\end{equation}
where $L$ is a Lorentz transformation, satisfying
$L^\transpose \! \eta L = \eta$.
The right hand side of equation~(\ref{eLaU})
is a product of the $4 \times 4$ matrix $L$,
the $4 \times 3$ matrix
$\left( \begin{array}{c} 0 \\ a \end{array} \right)$,
and the $3 \times 3$ matrix $U$.
The Lorentz transformation $L$ contains
the 6 redundant Lorentz degrees of freedom
of the spatial dynamical vierbein.
In BSSN gauge, where $\tilde{e}_{0\alpha} = 0$,
the Lorentz transformation $L$ reduces to a purely spatial rotation.

The choice of whether $U$ is upper or lower triangular,
or orthogonal, has no effect on the numerical integration;
the decomposition is needed only to project out the physical
degrees of freedom.
The choice adopted in this paper is $U$ upper triangular.
The 6 physical degrees of freedom of the gravitational field
are then the 3 components $a_a$ of the spatial diagonal matrix $a$,
and the 3 above-diagonal components $U_{ab}$
of the upper-triangular matrix $U$,
\begin{equation}
\label{aUdyn}
  a
  =
  \left(
  \begin{array}{ccc}
  a_1 & 0 & 0 \\
  0 & a_2 & 0 \\
  0 & 0 & a_3
  \end{array}
  \right)
  \ , \quad
  U
  =
  \left(
  \begin{array}{ccc}
  1 & U_{12} & U_{13} \\
  0 & 1 & U_{23} \\
  0 & 0 & 1
  \end{array}
  \right)
  \ .
\end{equation}


\kninfBHfourtwoafig

\kninfBHfourtwoqfig

\section{Results}

\subsection{Parameters}

The conformally separable solutions hold in the asymptotic limit of small
(but nonzero) outgoing and ingoing accretion rates.
But the smaller the accretion rates,
the more rapidly inflation exponentiates,
and the sooner the integration stalls because the
singular-value matrix $A_{\tform}$ in equation~(\ref{ApiTC})
becomes ill-conditioned.
The accretion rates adopted in this paper
are a compromise,
\begin{equation}
\label{vuparameters}
  \vel = 0.01
  \ , \quad
  \uel = 0.02
  \ ,
\end{equation}
small enough that the conformally separable solution is a satisfactory
approximation
(a proposition that can be tested by the degree to which the
constraint equations are satisfied),
but large enough that the numerical integration continues
through several BKL bounces.
The outgoing and ingoing accretion rates in the initial conditions are,
equation~(\ref{Npm}),
proportional to $\uel \pm \vel$,
\begin{equation}
  N^{\pm} \propto \uel \pm \vel
  \ .
\end{equation}

The black hole spin adopted in this paper is
\begin{equation}
\label{aparameter}
  a = 0.96
  \ ,
\end{equation}
which is chosen to be large, but short of extremal.
As noted by \cite{Hamilton:2010b},
the conformally separable solutions do not admit an extremal black hole,
since they require that the inner horizon be separate from the outer horizon.

Numerical experiment indicates that the most ``difficult'' computation
(in the sense that the Hamiltonian constraints are least well satisfied)
is at mid latitudes, where gradients in the latitude direction are greatest.
To illustrate the results, this paper therefore adopts
\begin{equation}
  \mbox{latitude} = 42^\circ
  \ .
\end{equation}
Results at a variety of latitudes are shown later,
in \S\ref{latitudes-sec}.

\subsection{BKL behavior}

Figure~\ref{kninfBH42a}
shows the lengths $a_a$ and spatial rotation $U_{ab}$,
equations~(\ref{aUdyn}),
of the 3 axes of the dynamical spatial vierbein
as a function of the mean scale factor
$\bar{a} \equiv ( a_1 a_2 a_3 )^{1/3}$.
The parameters are those of equations~(\ref{vuparameters})
and~(\ref{aparameter}).
The Figure shows results
computed using respectively the
BSSN and principal gauges.
The two gauges yield similar but slightly different results.

The behavior
illustrated in Figure~\ref{kninfBH42a}
is characteristic of BKL collapse
(see \S\ref{BKL-sec}).
The evolution encompasses 4 Kasner epochs
during which the scale factors $a_a$ evolve as powers
of the mean scale factor $\bar{a}$,
punctuated by 3 BKL bounces
during which the power-law behavior changes abruptly,
and at the same time the rotation of the axes changes abruptly.

Figure~\ref{kninfBH42q}
shows the Kasner exponents $q_a$ inferred
from the evolution of the scale factors
shown in Figure~\ref{kninfBH42a},
\begin{equation}
  a_a \propto \bar{a}^{3 q_a}
  \ .
\end{equation}
By definition of the mean scale factor
$\bar{a} \equiv ( a_1 a_2 a_3 )^{1/3}$,
the sum of the exponents is unity, $\sum q_a = 1$.
Figure~\ref{kninfBH42q}
shows in addition the sum $\sum q_a^2$ of the squares
of the Kasner exponents,
which the BKL model predicts should also equal unity,
equations~(\ref{kasnerexponents}).
The numerically calculated Kasner exponents $q_a$
generally conform to the BKL model during Kasner epochs,
but deviate during BKL bounces.

The dashed lines shown in Figures~\ref{kninfBH42a} and~\ref{kninfBH42q}
at small mean scale factor $\bar{a}$
are the predictions of the conformally separable solution,
continued up to the point where the conformally separable solution
is predicted to fail because of growing rotational motions
\cite{Hamilton:2010a,Hamilton:2010b}.
The numerically computed results
are indeed consistent with the predictions of the conformally separable solution
approximately up to the expected point of failure.

\kninfBHfourtwoHfig

\subsection{Hamiltonian constraint}
\label{Ham-sec}

A figure of merit for how well the numerical computation is
performing is provided by the Hamiltonian constraint,
which is the all-spatial component ($123\alpha\beta\gamma$)
of the difference between the curvature $\bPi$
and ($\kappa \equiv 8\pi G$ times)
the (modified) energy-momentum tensor $\tilde{\bT}$,
and which should equal zero if the computation is accurate,
\begin{equation}
\label{Ham}
  ( \Pi
  -
  \kappa
  \tilde{T} )_{123\alpha\beta\gamma}
  \approx 0
  \ .
\end{equation}
There are of course other constraints;
but the Hamiltonian constraint~(\ref{Ham}) is symptomatic.

\kninfBfourtwoDrfig

Figure~\ref{kninfBH42H}
shows the Hamiltonian constraint~(\ref{Ham})
for the same model as in
Figures~\ref{kninfBH42a} and~\ref{kninfBH42q},
for the BSSN and principal gauges.
Figure~\ref{kninfBH42H} also shows for comparison
the (modified) energy-momentum tensor $\tilde{T}_{123\alpha\beta\gamma}$,
and the potential energy term
$\left( \tfrac{1}{4} \be \wedgie [ \bGamma , \bGamma ] \right)_{123\alpha\beta\gamma}$
in equation~(\ref{Einsteineqaltformd}).
Figure~\ref{kninfBH42H}
shows that BSSN gauge performs better than principal gauge,
in the sense that the Hamiltonian constraint is relatively smaller than the
the energy-momentum.

The Hamiltonian constraint shown in
Figure~\ref{kninfBH42H}
is not zero either in the initial conditions or subsequently.
The reason the Hamiltonian constraint is not zero
in the initial conditions
is that the initial conditions are taken to be those of the
conformally separable solution,
which is not exact for finite accretion rates.
The Hamiltonian constraint gets worse as the integration proceeds because,
as described in \S\S\ref{singledirection-sec} and~\ref{factorvierbein-sec},
the integration proceeds along a single radial direction at fixed latitude,
with gradients in angular directions being taken from the conformally separable
solution as opposed to being computed numerically.
Thus the growing departure of the Hamiltonian constraint from zero
can be attributed at least in part to
a failure of the conformally separable solution to yield a good
approximation to angular gradients.

Numerical experiment shows that, up to a certain point,
the closer to the inner horizon the numerical integration is started,
the better the Hamiltonian and Gaussian constraints are satisfied
in the conformally separable initial conditions.
The constraints are best satisfied when the integration is started
around the end of inflation,
at approximately the boundary between the first two Kasner epochs.
This accounts for the location of the starting point
of the numerical integration indicated by the solid lines in
Figures~\ref{kninfBH42a} and~\ref{kninfBH42q}.
In practice, the radius at which the integration is started is
\begin{equation}
  r_{\rm init}
  =
  0.72002 \, M
  \ ,
\end{equation}
which is $r_{\rm init} - r_- = 0.00002 \, M$
above the radius $r_- = 0.72 \, M$ of the inner horizon
for spin parameter $a$ given by equation~(\ref{aparameter}).
If the integration is started at larger radius,
then the Hamiltonian constraints are less well satisfied,
and the integration breaks down earlier than shown in
Figures~\ref{kninfBH42a} and~\ref{kninfBH42q}.

I experimented, unsuccessfully,
with adjusting the initial conditions in an attempt to improve
the Hamiltonian constraints one way or another.
The conformally separable solutions are approximate, not exact,
and within the scope of these solutions
it is impossible to satisfy all constraints
(Hamiltonian and Gaussian) exactly.

In Principal gauge,
the Hamiltonian constraint is least well satisfied
around $\bar{a} \sim 10^{-3}$.
This coincides with the proper acceleration $\Gamma_{a00}$
of the tetrad frame becoming large,
especially in the $a = 2$ latitude direction
(large compared for example to the acceleration in BSSN gauge).
A related circumstance is that in BSSN gauge
also around $\bar{a} \sim 10^{-3}$,
the principal null directions cease to point
mainly along the tetrad 1-direction,
as seen in Figure~\ref{kninfBH42mom},
the direction along which the null directions are forced to point
in Principal gauge.

\kninfBHfourtwomomfig

\subsection{Radius and horizon function}

Figure~\ref{kninfB42Dr}
shows the radius $r$ and horizon function $\Delta_r$
in the same model as in
Figures~\ref{kninfBH42a} and~\ref{kninfBH42q}.
The evolution of the radius and horizon function
are governed by equations~(\ref{dr}) and~(\ref{dDelta}).

The radius $r$ plays a role in setting up
the conformally separable initial conditions,
but, as alluded to after equation~(\ref{daa}),
thereafter does not enter directly into the either
the equations of motion~(\ref{eqaltformd})
of the gravitational field
or~(\ref{dp}) and~(\ref{dn}) of the collisionless streams.
Rather, the radius $r$ is needed
only to determine angular derivatives of the fixed vierbein
$\mathring{e}^\kappa{}_\mu$,
equation~(\ref{kerrvierbein})
(specifically,
the latitude gradient $\partial / \partial \theta$
of the separable conformal factor $\rhosep$, equation~(\ref{rhosep}),
depends on $r$).
In other words, the behavior of the radius $r$
is not a central element of the evolution.
Figure~\ref{kninfB42Dr}
serves merely to demonstrate that $r$ does not go crazy.

Similarly, the horizon function $\Delta_r$
plays a role in setting up
the conformally separable initial conditions,
but after the initial conditions are set,
nothing in the equations of motion
of either the gravitational field or the collisionless streams
depends on the horizon function.
Figure~\ref{kninfB42Dr}
serves merely to demonstrate that the horizon function $\Delta_r$
``freezes'' following inflation,
as commented after equation~(\ref{Deltarth}).

\subsection{Collisionless streams}

Figure~\ref{kninfBH42mom}
shows the spatial components $p^\pm_a$ of the tetrad-frame momenta,
and densities $N^\pm$,
of the outgoing and ingoing collisionless null streams
that provide the source of energy-momentum in the model.
Also shown are the momenta and densities of principal outgoing
and ingoing null streams, designated {\scriptsize (PN)}.
There is no actual energy-momentum in the principal streams;
the behavior of the densities $N^\pm \mbox{\scriptsize{(PN)}}$
is shown only for comparison.
The collisionless streams that constitute the source of energy-momentum
are initially almost aligned with the principal null directions,
but then deviate.

Principal gauge is defined by the requirement that the
tetrad frame remains aligned with the principal null directions,
\S\ref{princ-sec},
the principal null momenta satisfying
$p^+_1 = - p^-_1$
and $p^\pm_2 = p^\pm_3 = 0$.
The top-right panel of Figure~\ref{kninfBH42mom}
confirms that the principal null momenta indeed satisfy these conditions.

\kninfBafig

\kninfBqfig


\subsection{Results at various latitudes}
\label{latitudes-sec}

Figure~\ref{kninfBa}
shows the evolution of the scale factors $a_a$
at a variety of latitudes,
for the same accretion rates~(\ref{vuparameters})
and black hole spin~(\ref{aparameter}) as before.
The gauge is BSSN; results in principal gauge are similar
but not identical.
Results are shown for latitudes from the equator
almost to the pole.
The highest latitude, $89^\circ$,
is not quite at the pole
in order to avoid the coordinate singularity at the pole
that occurs when polar coordinates $\theta$, $\phi$ are used, as here.
In polar coordinates,
the determinant of the vierbein~(\ref{kerrvierbein}) is zero at the pole,
and the inverse vierbein is singular.
The singularity could presumably be removed by using different coordinates,
but that is not done in this paper.

Figure~\ref{kninfBa}
shows BKL-like behavior at all latitudes.
At all latitudes except the equator, $0^\circ$,
the numerical integration continues over 4 Kasner epochs
and 3 BKL bounces before grinding to a halt.
At the equator,
the integration covers just 2 Kasner epochs and 1 BKL bounce
before stalling.

Figure~\ref{kninfBq}
shows the Kasner exponents $q_a$
corresponding to the models shown in
Figure~\ref{kninfBa}.
Again,
the behavior is consistent with the BKL model,
where the exponents $q_a$ satisfy the Kasner relations~(\ref{kasnerexponents})
during power-law Kasner epochs.

\section{Summary and conclusions}

The inner horizons of accreting, rotating black holes are
subject to the mass inflation instability discovered by
Poisson \& Israel
\cite{Poisson:1989zz,Poisson:1990eh,Barrabes:1990}.
During inflation,
collisionless accretion streams
focus along the outgoing and ingoing principal null directions,
their momenta and densities growing exponentially huge.

This paper has explored numerically the behavior of
collisionless outgoing and ingoing accretion streams,
and of all 6 physical degrees of freedom of the gravitational field,
during and following inflation
in the vicinity of the inner horizon of an accreting, rotating black hole.
The generic outcome appears to be BKL collapse,
as originally proposed in the 1970s by
Belinskii, Khalatnikov, and Lifshitz
\cite{Belinskii:1970,Belinskii:1971,Belinskii:1972,Belinskii:1982}.
The result differs from the common claim
that the generic outcome of inflation is
a weak null singularity on the Cauchy horizon
\cite{Ori:1992,Ori:2001pc,Dafermos:2012ny,Luk:2013cqa}.
As argued by \cite{Hamilton:2008zz},
the outcome of a weak null singularity is an artifact of the assumption
that a black hole remains isolated for ever after it first collapses,
whereas real astronomical black holes are never isolated,
but rather accrete.

Part of the aim of this paper was to test numerically the
conformally separable solutions for accreting, rotating black holes
discovered by
\cite{Hamilton:2010a,Hamilton:2010b}.
The conformally separable solutions are not exact,
but are claimed by
\cite{Hamilton:2010a,Hamilton:2010b}
to hold asymptotically in the limit of small but nonzero accretion rates.
In the conformally separable solutions, inflation is followed by collapse,
but the solutions are expected
to break down at small scales when rotational motions become important.
The present paper takes the conformally separable solutions
as initial conditions for numerical integration.
The numerical computations confirm that the conformally separable solutions
are valid over the expected range of validity.
In fact, the conformally separable solutions appear to describe the first two
Kasner epochs of BKL collapse.
The numerical integrations in the present paper continue through 4 Kasner epochs
punctuated by 3 BKL bounces,
before the integrations grind to a halt.

The numerical computations use a covariant Hamiltonian tetrad approach
devised by the author
\cite{Hamilton:2016mmc}.
Computations have been carried out in two gauges,
principal gauge, in which the tetrad is aligned with the principal null directions,
and BSSN (strictly, BSSN-like) gauge,
which imposes the traditional ADM gauge choice~(\ref{admgauge}).
The two gauges yield similar results,
although BSSN appears to be superior in the sense that
the Hamiltonian constraints are better satisfied by BSSN.

A major simplifying assumption made in this paper,
\S\ref{singledirection-sec},
is to assume that gradients in spatial directions
$t$, $\theta$, $\phi$
continue to be given by the conformally separable solution
even after the latter breaks down.
The assumption is motivated in part by the BKL argument that
spatial gradients are subdominant to gradients in the time direction
during BKL collapse.
The simplification allows the numerical integration to be carried inward
into the black hole along a single radial direction at a single latitude.
The approximation may hold approximately in the conformally separable regime
(first two Kasner epochs),
but certainly fails thereafter,
as evidenced both by the growing failure of the Hamiltonian constraint,
Figure~\ref{kninfBH42H},
and by the differences in evolution at different latitudes,
Figure~\ref{kninfBa}.

Full exploration of the outcome of the mass inflation instability
in accreting, rotating black holes
will require calculating spatial gradients in full numerical general relativity.
This will be a challenging task.


\begin{acknowledgments}
This research was supported in part by FQXI mini-grant FQXI-MGB-1626.
\end{acknowledgments}

\section*{References}

\bibliography{bh}

\end{document}